# How does Mg$^{2+}_{(aq)}$ interact with ATP$_{(aq)}$? Biomolecular structure through the lens of liquid-jet photoemission spectroscopy

Karen Mudryk,[1,‡] Chin Lee,[1,2,3,†,‡] Lukáš Tomaník,[4,§,‡] Sebastian Malerz,[1] Florian Trinter,[1,5] Uwe Hergenhahn,[1] Daniel M. Neumark,[2,3] Petr Slavíček,[4]* Stephen Bradforth,[6]* and Bernd Winter[1]*

[1]Fritz-Haber-Institut der Max-Planck-Gesellschaft, Faradayweg 4-6, 14195 Berlin, Germany

[2]Department of Chemistry, University of California, Berkeley, CA 94720, USA

[3]Chemical Sciences Division, Lawrence Berkeley National Laboratory, Berkeley, CA 94720, USA

[4]Department of Physical Chemistry, University of Chemistry and Technology, Prague, Technická 5, Prague 6 16628, Czech Republic

[5]Institut für Kernphysik, Goethe-Universität Frankfurt, Max-von-Laue-Straße 1, 60438 Frankfurt am Main, Germany

[6]Department of Chemistry, University of Southern California, Los Angeles, CA 90089, USA

**ABSTRACT:** Liquid-jet photoemission spectroscopy (LJ-PES) allows for a direct probing of electronic structure in aqueous solutions. We show the applicability of the approach to biomolecules in a complex environment, exploring site-specific information on the interaction of adenosine triphosphate in the aqueous phase (ATP$_{(aq)}$) with magnesium (Mg$^{2+}_{(aq)}$), highlighting the synergy brought about by the simultaneous analysis of different regions in the photoelectron spectrum. In particular, we demonstrate intermolecular Coulombic decay (ICD) spectroscopy as a new and powerful addition to the arsenal of techniques for biomolecular structure investigation. We apply LJ-PES assisted by electronic-structure calculations to study ATP$_{(aq)}$ solutions with and without dissolved Mg$^{2+}$. Valence photoelectron data reveal spectral changes in the phosphate and adenine features of ATP$_{(aq)}$ due to interactions with the divalent cation. Chemical shifts in Mg 2p, Mg 2s, P 2p, and P 2s core-level spectra as a function of the Mg$^{2+}$/ATP concentration ratio are correlated to the formation of [Mg(ATP)$_2$]$^{6-}_{(aq)}$, [MgATP]$^{2-}_{(aq)}$, and [Mg$_2$ATP]$_{(aq)}$ complexes, demonstrating the element-sensitivity of the technique to Mg$^{2+}$–phosphate interactions. The most direct probe of the intermolecular interactions between ATP$_{(aq)}$ and Mg$^{2+}_{(aq)}$ is delivered by the emerging ICD electrons following ionization of Mg 1s electrons. ICD spectra are shown to sensitively probe ligand exchange in the Mg$^{2+}$–ATP$_{(aq)}$ coordination environment. In addition, we report and compare P 2s data from ATP$_{(aq)}$ and adenosine mono- and di-phosphate (AMP$_{(aq)}$ and ADP$_{(aq)}$, respectively) solutions, probing the electronic structure of the phosphate chain and the local environment of individual phosphate units in ATP$_{(aq)}$. Our results provide a comprehensive view of the electronic structure of ATP$_{(aq)}$ and Mg$^{2+}$–ATP$_{(aq)}$ complexes relevant to phosphorylation and dephosphorylation reactions that are central to bioenergetics in living organisms.

## INTRODUCTION

Photoemission spectroscopy is a method of choice for the characterization of solid-state systems as it directly probes the electronic structure of materials and surfaces.[1] The application of this powerful tool to aqueous solutions – the main domain of chemistry – has been complicated by the high water vapor pressure inherent to these systems, which had prevented the detection of the photo-emitted electrons. However, the advent of the liquid microjet technology marked a breakthrough.[2] This advance allows high-quality photoemission spectra to be recorded on an absolute energy scale[3], providing novel chemical insight into a number of areas, *e.g.*, the electronic structure of liquid water and coordination compounds,[4, 5] ion pairing,[6] acid-base properties in poly-protic systems,[7] oxidation of nucleic acids,[8] or the depth profile of air–water interfaces.[9] The technique has been so far rarely used in the field of biochemistry, which includes large molecules in complex environments. Here, we demonstrate how liquid-jet photoemission spectroscopy (LJ-PES) assesses specific molecular interactions that can provide structural information for such systems, exemplified here for the buffered adenosine triphosphate nucleotide (ATP$_{(aq)}$).

The interaction of molecules with soft X-ray photons leads to the photoemission of electrons generated in different processes. Let us begin by inspecting the possible spectral regions that can be recorded in a LJ-PES measurement;

see Figure 1(a). The valence electrons are the easiest to ionize; their binding energies (BEs) are observed in the 10–20 eV range for virtually all solvated molecules, including water.[4, 10] This part of the photoemission spectrum bears, therefore, rather limited structural information but it is important for understanding, *e.g.*, redox properties.[8] The core-level electrons (1s) are often used as a sensitive probe of the molecular constitution.[7] They are energetically well-separated for different atoms, and in addition, their chemical environment controls the so-called "chemical shift". This spectroscopy is thus analogous to nuclear magnetic resonance (NMR) spectroscopy. The inner-valence photoelectrons (e.g., 2s or 2p) can bear similar type of information, especially for heavier atoms, but this spectral range is less explored. Note that within a single measurement we can obtain spectra associated to all atoms and all energy levels, provided a sufficiently high photon energy is used. We show below that photoelectron (PE) spectroscopy is applicable for disentangling even intermolecular interactions, although the chemical shifts brought about by these interactions are often surprisingly small.[11]

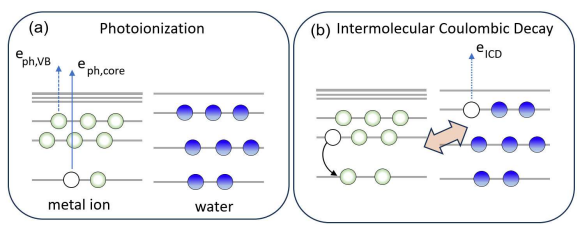

**Figure 1.** (a) Emitted valence or core-level photoelectrons upon photoionization of the ionic solute in water. (b) ICD: A deep core hole relaxes via refill by a valence electron from the same atom, and the released energy is used to ionize the first solvation shell; here depicted for a hydrating water molecule.

Yet another type of structural information is revealed by second-order electron emission, i.e., the Auger-type electrons. The core hole formed upon ionization can be refilled by an inner-shell valence electron, with the released energy causing autoionization of another valence electron. This is the Auger–Meitner process if it takes place within a single atom, and these electrons are only little affected by chemical environment.

In recent years, much attention has been paid to an analogous type of processes taking place intermolecularly. In intermolecular Coulombic decay (ICD),[12] the generated core hole is filled by a valence electron, and the excess energy is used to ionize a neighboring molecule; this non-local autoionization process is depicted in Figure 1(b). The probability of the process scales with $1/R^6$, where R is the distance between the two neighboring atomic or molecular entities. Hence, ICD uniquely probes the first solvation shell, and, equivalently, ion pairing and weak ion binding / association; the process is even sensitive to molecular orientation.[13] The energy of a specific ICD feature, associated with two locally separated valence holes (open circles in Figure 1(b)), depends on the electronic structure of each entity. ICD can serve as a molecular ruler, in analogy to the fluorescence resonance energy transfer (FRET) process.[14] In fact, both processes are controlled by the same Coulombic matrix elements, yet ICD is not subject to any strict selection rule.[12, 14] While this application of the ICD process has been hypothesized since its discovery,[12] the present work shows ICD's practical realization for the first time.

$ATP_{(aq)}$ consists of a nucleoside (adenosine, which is formed by adenine and ribose) bound to a chain of three phosphate groups [15] that enables energy exchange and signal transduction in living organisms.[16-21] ATP's molecular structure is illustrated in Figure 2, highlighting the adenine, ribose, and phosphate units. The latter are designated as α, β, and γ (where α refers to the phosphate directly bound to the nucleoside, β refers to the bridging phosphate unit, and γ refers to the terminal phosphate) and are shown in their deprotonated form ($ATP^{4-}_{(aq)}$, as predominantly found at physiological pH).[22] We also include $Mg^{2+}_{(aq)}$, or more specifically, $[Mg(H_2O)_6]^{2+}_{(aq)}$, the octahedral complex configuration adopted by the free ion in aqueous solution.[23] This ion interacts with one or more phosphate units to form various $Mg^{2+}$–$ATP_{(aq)}$ complexes,[24-26] such as $[Mg(ATP)_2]^{6-}_{(aq)}$,[24, 27-30] $Mg_2ATP_{(aq)}$,[31] and $[Mg_2(ATP)_2]^{4-}_{(aq)}$.[24, 32] For $[MgATP]^{2-}_{(aq)}$, a closed form has also been proposed, where the metal ion interacts not only with oxygene.g. atoms from the phosphate units but also, presumably through a water molecule, with a N atom from the adenine unit. The 'closed form' / 'open form' ratio was reported to be 1:10 in aqueous solution.[21, 33-36] A schematic representation of this configuration is shown in Figure S1 in the Supporting Information (SI).

The biological function of $ATP_{(aq)}$ stems from the liberation and addition of phosphate units, termed dephosphorylation and phosphorylation reactions, respectively, in the aqueous environment of the cell. [37] In the former case, energy is released *via* hydrolysis to produce adenosine diphosphate (ADP), and subsequent hydration of the latter. In the latter case, phosphorylation of $ADP_{(aq)}$ takes place to restore $ATP_{(aq)}$ in the cell. In this way, metabolic pathways involving $ATP_{(aq)}$ and $ADP_{(aq)}$ are regulated by chemical bond breaking and bond formation at the phosphate chain.[38] Despite a single phosphate unit seemingly being involved in phosphorylation and dephosphorylation reactions,[39] $Mg^{2+}_{(aq)}$ complexation to multiple sites in the phosphate chain is required for the overall reaction to proceed.[40-43] In the absence of enzymes, α-, β-, and γ-phosphate are expected to be equally involved in $Mg^{2+}_{(aq)}$ association to $ATP_{(aq)}$.[26, 44, 45] In enzyme-bound $ATP_{(aq)}$, the β- and γ-phosphate units are most likely to interact with $Mg^{2+}_{(aq)}$.[46] The binding constants associated with specific $Mg^{2+}$–$ATP_{(aq)}$ ion pairing motifs alter the Gibbs free energy of hydrolysis,[47] as they determine the concentration of each of the species involved in the reaction.

Overall, phosphorylation and dephosphorylation processes are strongly affected by the presence of divalent metal cations, particularly by $Mg^{2+}_{(aq)}$, the cation with the highest $ATP_{(aq)}$ binding affinity,[25, 34] and by hydration effects inherent to the aqueous environment of the cell.[46, 48-50] In this way, metal–ligand coordination at the phosphate chain is involved in dephosphorylation (hydrolysis) or phosphorylation at specific phosphate units *via* charge redistribution and conformational changes, [24, 46] with the relatively small size of the $Mg^{2+}_{(aq)}$ ion facilitating coordination. [51] These bonding interactions should result in charge redistribution at the P–O–P bond[39] and differences in electron BEs of the α-, β-, and γ-phosphates in $ATP_{(aq)}$ and $Mg^{2+}$–$ATP_{(aq)}$. Such intramolecular (and intermolecular) charge redistributions, as well as solvation effects, are reflected in the phosphorylation and dephosphorylation reaction mechanisms.[39, 40, 43, 52]. The nucleobase–metal-ion interactions mentioned earlier may also play a role in the overall reaction.[53] A large amount of association equilibria data has been determined in support of such effects.[54] However, the characterization of associated structures by advanced spectroscopic methods is lacking.

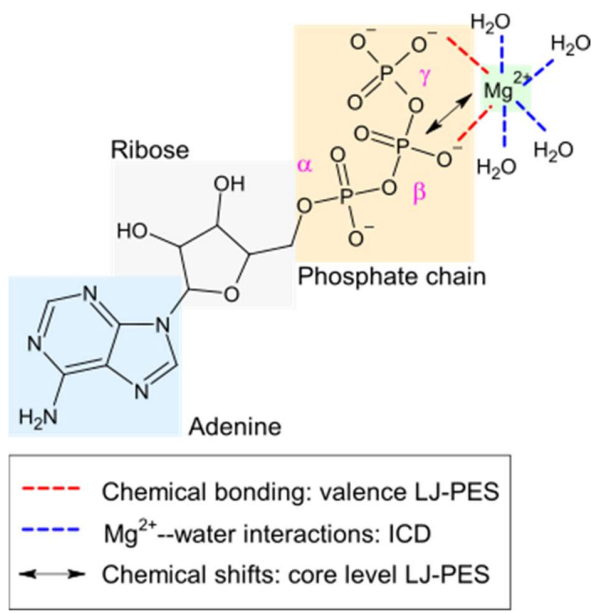

**Figure 2.** (a) Molecular structure of $ATP_{(aq)}$ in the deprotonated form ($ATP^{4-}_{(aq)}$), as predominantly found at physiological pH. Adenine, ribose, and α-, β-, and γ-phosphate units are labelled; one of many motifs for $Mg^{2+}$ binding is shown (see text). Molecular interactions probed using valence and core LJ-PES, as well as ICD spectroscopy, are highlighted using red and blue dashed lines, respectively.

With that in mind, we aim to investigate whether LJ-PES, [10, 55] which has been used to probe chemical shifts in inorganic and organic solutes in aqueous environments[7, 8, 56-58] is (1) sufficiently sensitive to $ATP_{(aq)}$ ion pairing and/or complexation, and (2) capable of distinguishing between the three phosphate groups. Regarding the first question, we report on the electronic-structure changes upon the addition of $Mg^{2+}$ to $ATP_{(aq)}$ solutions and the formation of different $Mg^{2+}$–$ATP_{(aq)}$ complexes, inferred from valence and Mg 2p, Mg 2s, P 2p, and P 2s core-level PE spectra. To elaborate on the second, we also present P 2s spectra from aqueous solutions of neat $ATP_{(aq)}$, $ADP_{(aq)}$, and $AMP_{(aq)}$, determining relative changes in the BEs of α-, β-, and γ-phosphate.

We contrast the (direct) PE spectra (compare Figure 1(a)) with non-local autoionization electron signal, upon ICD [5, 13, 14, 59, 60] of $Mg^{2+}_{(aq)}$ ions (compare Figure 1(b)) in the presence of $ATP_{(aq)}$ in their coordination environment. The unique sensitivity to the first hydration shell has been recently demonstrated for $Mg^{2+}$ in water [13] which lays the groundwork for the present study. The detected ICD electron signal senses the interaction of the metal ion with its immediate neighbor in an exclusive and most direct way, namely by the selective autoionization of just the constituent of the first solvation shell. To be more specific, we measure the ICD electrons formed upon photoionization of Mg 1s core-level electrons. Here, the respective core hole is refilled by electrons from the Mg 2s (or Mg 2p) core levels, and the released excess energy is used to ionize the surrounding molecules in the first solvation shell. The emitted second-order electron is then detected as an ICD signal; to be detailed below.

Ideally, the experiments could be interpreted without the need of *ab initio* theory. The changes of both the PE and ICD signals can indeed be generally interpreted within simple electrostatic concepts. However, these views can be oversimplified, and it is, therefore, still desirable to confirm the experiments with *ab initio* calculations. These calculations help to understand the quantitative correlations in the experiments.

Due to the importance of $ATP_{(aq)}$ in biochemistry, the $Mg^{2+}$–phosphate interactions have been previously explored by many techniques, *e.g.*, using aqueous-phase X-ray emission,[61] infrared,[36] Raman,[62] and nuclear magnetic resonance (NMR)[63] spectroscopies. The page $Mg^{2+}$–ATP coordination chemistry has been recently investigated in the gas phase using mass spectrometry (in particular, phosphate–$Mg^{2+}$–adenine interactions)[35] and in acetate solutions[64] using NMR spectroscopy. Previous LJ-PES work focused on the ribose or adenine units in $ATP_{(aq)}$[65-67], or on inorganic phosphate [6] aqueous solutions. The present LJ-PES results can thus be compared to these techniques, demonstrating its scope of applicability.

METHODS

**Experiments.** 0.5 M $ATP_{(aq)}$, $ADP_{(aq)}$, and $AMP_{(aq)}$ solutions were prepared by dissolving the required amount of adenosine 5'-triphosphate disodium salt hydrate, adenosine 5'-diphosphate acid, and adenosine 5'-monophosphate disodium salt (Carbosynth, 95%) in Millipore water, respectively. $ATP_{(aq)}$ and $ADP_{(aq)}$ samples containing $Mg^{2+}$ were prepared by addition of $Mg(NO_3)_2$ (ACROS ORGANICS, 99+%) to 0.5 M $ATP_{(aq)}$ or $ADP_{(aq)}$ solutions in order to reach 0.25:1, 0.5:1, 0.75:1, 1:1, and 1.5:1 $Mg^{2+}$/ATP concentration ratios. For each sample, the solution pH was adjusted to 8.2 by the addition of the required amount of Tris (tris(hydroxymethyl)aminomethane, Sigma Aldrich, ≥99.8%), to ensure that the $ATP_{(aq)}$ phosphate chain was fully deprotonated ($ATP^{4-}_{(aq)}$, see Figure 2).[22, 27, 68] This alternative approach to the traditional Tris/TrisHCl buffer pH adjustment methodology allows us to reduce the number of chem-ical species in the solution that can interfere with the observation of the PE signals of interest. The specific concentration of Tris$_{(aq)}$ in each sample is listed in Table S1 in the SI. The Mg 2s core-level PE spectra from a 0.5 M Mg(NO$_3$)$_{2(aq)}$ solution with a pH adjusted to 8.4 was measured for reference.

LJ-PES experiments were performed at the P04 beamline at PETRA III[69] (DESY, Hamburg, Germany) using the EASI setup.[70] Photoelectrons emitted from the sample were detected using a differentially pumped hemispherical electron analyzer at 130° with respect to the light propagation axis (circular polarization). The samples were delivered into the vacuum chamber of the EASI setup in the form of liquid microjets[71] using a glass capillary of 28 μm inner diameter and flow rates in the range of 0.55-0.80 mL/min. The sample temperature was kept at 10°C by means of a cooling system interfaced with the liquid-jet holder. A small metallic wire was placed into the main polyether ether ketone (PEEK) liquid delivery line to electrically connect and ground the liquid jet to EASI. The liquid jet's (horizontal) flow axis was perpendicular to both the light propagation (floor plane) and the electron detection (at an angle of 130° with respect to the photon beam) axes. PE spectra from ATP$_{(aq)}$ with dissolved Mg$^{2+}$ were recorded using a photon energy of 250 eV, spanning over the valence, Mg 2p, Mg 2s, P 2p, and P 2s spectral regions. To determine core-level chemical shifts with better precision, additional PE spectra focused on measuring only the Mg 2s, P 2p and P 2s regions with finer step width (0.05 instead of 0.2 eV) were recorded at 250 eV, 270 eV and 330 eV photon energy, respectively. Immediately before or after each of these a spectrum of the water valence region was measured with all other parameters identical, to aid in BE calibration (compare footnote §). P 2s PE spectra from AMP$_{(aq)}$, ADP$_{(aq)}$, and ATP$_{(aq)}$ were also recorded, using a photon energy of 330 eV. ICD experiments were performed at 1314 eV. The overall instrumental energy resolution was, approximately 210 meV at 250 eV, 220 meV at 330 eV, and 570 meV at 1314 eV photon energy.

The BE scale in the data presented here was calibrated based on the liquid water 1b$_1$ BE of 11.33 eV,[3] as commonly adopted in LJ-PES experiments.[11] While we have recently reported a more robust methodology for determining absolute aqueous-phase electron BEs,[3] (see footnote §) the data acquisition of the results presented here preceded those developments. However, absolute BEs are not the principal quantity of interest in this work, but rather relative spacings of peaks that can be attributed to different species in solution.

**Computations.** α-, β-, and γ-phosphate P 2s BEs of ATP$^{4-}$$_{(aq)}$, [Mg(ATP)$_2$]$^{6-}$$_{(aq)}$, [MgATP]$^{2-}$$_{(aq)}$, Mg$_2$ATP$_{(aq)}$, and [MgADP]$^{-}$$_{(aq)}$, were calculated using the maximum-overlap method (MOM) [72] as implemented in the Q-Chem 6.0 software[73] (the sample input for these calculations can be found in the SI). An excellent computational cost/performance ratio was previously demonstrated with this approach.[74] Due to the system size, we pragmatically employed the Hartree–Fock (HF) method with a core-enhanced aug-cc-pCVTZ basis set on P atoms and aug-cc-pVTZ basis set on other atoms. To model Mg 2s and 2p BEs, a core-enhanced aug-cc-pCVTZ basis set was used for Mg atoms and aug-cc-pVTZ for other atoms. The aqueous solution was modeled by the cluster–continuum approach.

For our quantum system, we explicitly included 26 water molecules around the triphosphate chain (to cover each terminal O with at least three hydrogen bonds from water molecules) of ATP$^{4-}$$_{(aq)}$, [MgATP]$^{2-}$$_{(aq)}$ and Mg$_2$ATP$_{(aq)}$, 17 water molecules for [MgADP]$^{-}$$_{(aq)}$, and 44 water molecules for [Mg(ATP)$_2$]$^{6-}$$_{(aq)}$ to screen the high charge density while the rest of the solvent was described by the polarizable-continuum model (PCM).[75, 76] We used the non-equilibrium variant of PCM with integral-equation formalism (IEF), Bondi radii, and recommended scaling factor α = 1.2.[77] Note that calculations of multiply charged ions are known to have limitations [6] as, *e.g.*, including extensive explicit solvation shell or counterions might be required. Consequently, our computed BEs will only provide a framework for the spectra interpretation rather than precise values. For pure Mg$^{2+}$$_{(aq)}$ solutions, the cation was solvated by six explicit water molecules in an octahedral geometry, while the rest of the solvent was modeled by PCM.

The calculated P 2s BEs were corrected for the missing electronic-correlation description in the HF method as follows. We used pyrophosphate (HP$_2$O$_7$$^{2-}$) hydrated by five explicit water molecules as a smaller model of the phosphate chain. We computed the P 2s BEs using the HF and the second-order Møller–Plesset (MP2) methods to approximately determine the error connected to the missing electronic correlation in the HF method. The correction was estimated to be +0.33 eV. The calculations of BEs were done on single structures optimized on the HF/6-31+G* level of theory with PCM (IEF, Bondi radii, α = 1.2) using Gaussian 09, revision D.01[78] (the Cartesian coordinates can be found in the SI). Due to its size, [Mg(ATP)$_2$]$^{6-}$$_{(aq)}$ with 44 water molecules could not be fully optimized and the structure obtained after 144 optimization steps was used.

The valence-region BEs of HOMO, HOMO−1, and HOMO−2 were calculated using ΔSCF and MOM, respectively, with CAM-B3LYP functional[79] and the aug-cc-pVTZ basis set. The solvation was modeled by including 5 explicit water molecules (three around the phosphate chain and two around the adenine moiety) and PCM (IEF, Bondi radii, α = 1.2). The calculations were performed for two optimized structures, one for the ATP molecule without the Mg ion and one corresponding to the "open form" of Mg-ATP. The optimization of structures was done on the CAM-B3LYP/6-31+G* level of theory, according to our previous work. [7] The Cartesian coordinates of the optimized structures can be found in the SI.

**RESULTS AND DISCUSSION**

**Mg$^{2+}$–ATP$_{(aq)}$ interaction: perspective of element-specific photoelectron spectra.** This section presents valence and core-level PE spectra from ATP$_{(aq)}$ samples with dissolved Mg$^{2+}$, providing an overview of the electronic structure of ATP$_{(aq)}$ and the interactions between the metal cation and different units in the ATP$_{(aq)}$ molecule.



PE spectra from 0.5 M $ATP_{(aq)}$ solutions with $Mg^{2+}_{(aq)}$ at different $Mg^{2+}$/ATP concentration ratios recorded at a photon energy of 250 eV are shown in Figure 3. The displayed wide energy range covers ionization of the valence band, and the Mg 2p, Mg 2s, P 2p, and P 2s core-levels. For reference, the respective spectrum recorded from 0.5 M $ATP_{(aq)}$ without $Mg^{2+}_{(aq)}$ is also shown. The as-measured spectra were energy calibrated (see Footnote §) as described in the Methods section, and intensities were normalized to yield the same peak height as that of the liquid water lowest ionization feature, $1b_1$. Spectral features from the $Na^+$ counter ion in the ATP salt used to prepare the solutions are labeled according to reference[80]. The intensity of this feature is observed to decrease as the $Mg^{2+}_{(aq)}$ concentration increases, which might be due to changing propensity of the solute under study at the surface. Contributions from the $NO_3^-_{(aq)}$ counter ion from the $Mg^{2+}$ salt used in the experiments are expected in the 9.0–9.5 eV BE range.[10] Contributions from Tris added to adjust the pH are labeled based on reference.[8] Valence PE data from $Tris_{(aq)}$ solutions is shown in Figure S2 in the SI, along with a discussion of its potential effect on spectral shifts. A zoom into the different spectral regions is shown in Figure 4 [panels (a)-(e)].

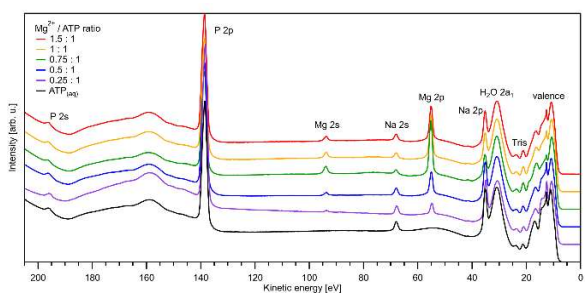

**Figure 3.** Valence as well as Mg 2p, Mg 2s, P 2p, and P 2s core-level PE spectra from $ATP_{(aq)}$ samples containing $Mg^{2+}_{(aq)}$ as a function of the $Mg^{2+}$/ATP concentration ratio, measured at 250 eV photon energy. The leading peaks are from ionization of the water $1b_1$ orbital. The bottom spectrum is from an $ATP_{(aq)}$ without added $Mg^{2+}_{(aq)}$.

Figure 4(a) expands the valence region, which is dominated by PE peaks arising from (direct) ionization of water valence electrons,[4] labeled as (liquid- and gas-phase) $1b_1$, $3a_1$, $1b_2$, and $2a_1$ (the latter only included in Figure 3). Observed small energy shifts and intensity variations of the $1b_{1(g)}$ gas-phase peak, from the surrounding water vapor, result from the combined effects of change of solution surface potential and small changes of water electronic structure upon addition of solute. [3, 81] Based on a previous valence LJ-PES study of $AMP_{(aq)}$ [8] and valence PE spectra from $ADP_{(aq)}$ and $ATP_{(aq)}$ (see Figure S3 in the SI for details), the phosphate, ribose, and adenine PE signatures contribute in the 8–10 eV BE range, with an additional adenine feature also present at 7–8 eV, as highlighted in the inset of Figure 4(a). The latter peak is observed to shift as the $Mg^{2+}$/ATP concentration ratio increases. Similar shifts are observed in the valence PE spectra recorded from $ADP_{(aq)}$ solutions as a function of the $Mg^{2+}$/ADP concentration ratio (see Figure S4 in the SI). These energy shifts for ATP are in qualitative agreement with our *ab initio* calculations. As shown in Table 1, the low-energy signal is comprised of the ionization of adenine's highest occupied molecular orbital (HOMO), as well as HOMO−1 and HOMO−2. It is seen that on average, the BEs are larger in the presence of $Mg^{2+}$.

**Table 1. Calculated Valence BEs (in eV) for ATP and ATP–Mg structures**

|  | HOMO | HOMO−1 | HOMO−2 |
|---|---|---|---|
| ATP without Mg | 7.47 | 7.69 | 7.99 |
| ATP–Mg | 7.40 | 8.18 | 8.43 |

Figure 4(a) reveals further $Mg^{2+}$-concentration-dependent spectral changes, in the 10–20 eV BE range, which can be also associated with the ionization of adenine and ribose orbitals, at lower BEs, and phosphate chain at higher energies.[82] Most noticeable is the emerging 20-eV BE peak (see the arrow) upon increasing the $Mg^{2+}$/ATP concentration ratio which can be attributed to the presence of $Mg^{2+}$–phosphate interactions in any of the $Mg^{2+}$–$ATP_{(aq)}$ moieties in which O atoms from the phosphate chain are directly exposed to the positive charge from the divalent metal cation. Previous LJ-PES experiments have been shown to be sensitive to $Na^+$–phosphate electrostatic interactions,[6] and we expect the presence of $Mg^{2+}_{(aq)}$ to have a more pronounced effect. To confirm our assignment, we have performed theoretical calculations employing the same methodology used to calculate P 2s BEs as detailed in the computations section. Our results indicate that the 20-eV BE feature originates from ionization of electrons from mixed, delocalized molecular orbitals appearing when $Mg^{2+}_{(aq)}$ ions are closely bound to the phosphate chain, with significant contributions from O, C, and P atoms. We note that O 2s signatures from phosphate compounds are expected in the 24.0–25.7 eV BE range [83], and are likely hidden underneath the water $2a_1$ peak. Another striking spectral change when increasing $Mg^{2+}$/ATP concentration ratio is the larger intensity on the low-BE side of the water $1b_1$ peak, occurring approximately between 9.5–11.0 eV BE. As we have shown in a previous work [8], this region contains contributions from phosphate, centered near 9 eV BE, and ribose contributions, around 10 eV.

More specific information on $Mg^{2+}$–ATP interaction is retrieved from the core-level spectra. For samples containing $Mg^{2+}_{(aq)}$, the Mg 2p and Mg 2s core-level photoelectron signal occurs near 55 eV and 95 eV, and is highlighted in Figures 4(b) and 4(c), respectively. With increasing $Mg^{2+}$/ATP concentration ratio, the respective intensities increase, and the peak centers shift to higher BEs; these shifts are indicated by dashed lines. We observe similar trends in the P 2p and P 2s spectral regions, near 140 and 196 eV BEs, as highlighted in Figures 4(d) and 4(e). All observed chemical shifts are plotted in Figure 5(a). These shifts can be qualitatively assigned to a change of the population of different $Mg^{2+}$-ATP species, as we detail later.



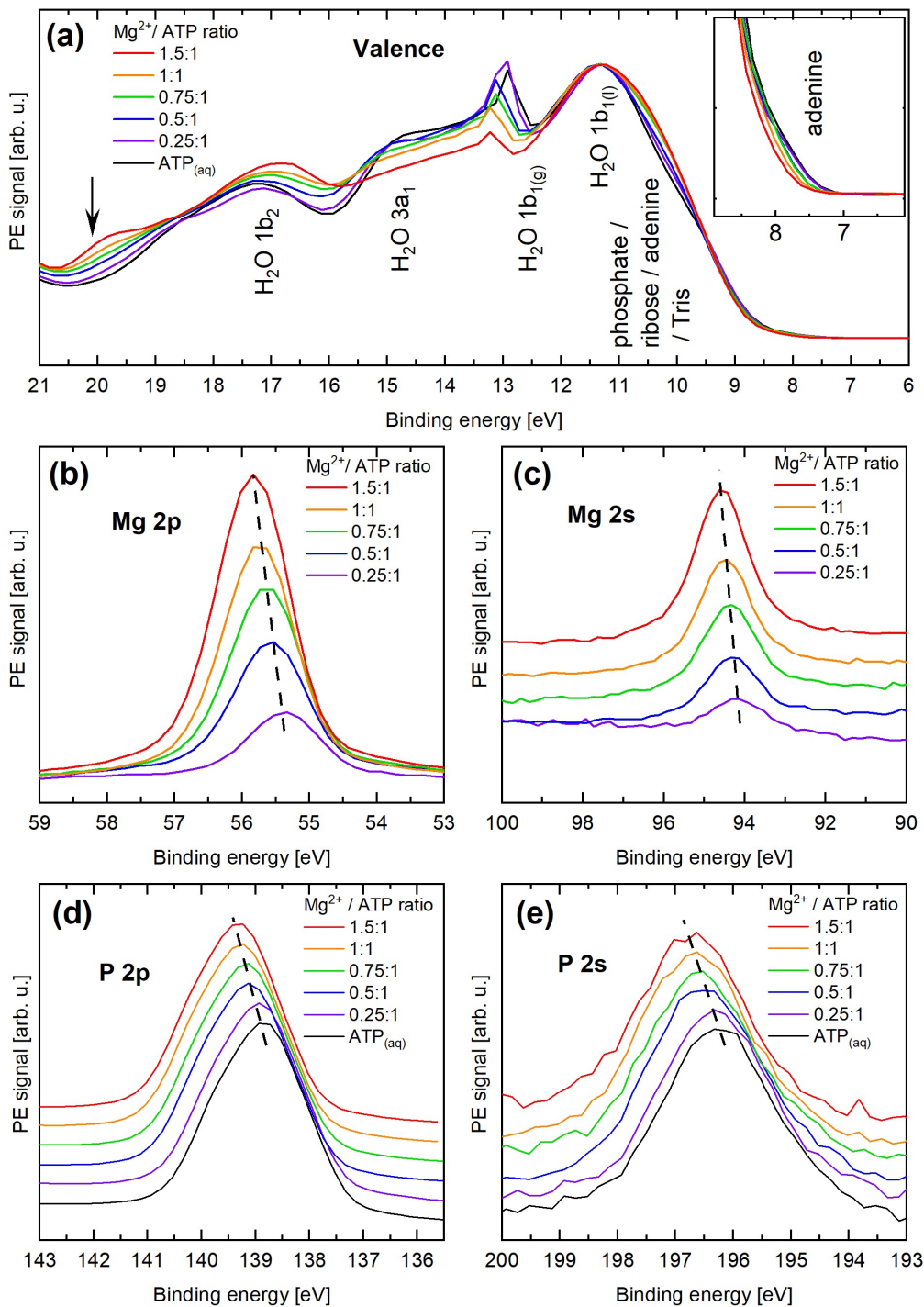

**Figure 4**. Highlights of the valence [panel (a)], Mg 2p [panel (b)], Mg 2s [panel (c)], P 2p [panel (d)], and P 2s [panel (e)] spectral regions from the data presented in Figure 3. In panel (a), PE signatures of ionization of water valence electrons are labelled as (liquid- and gas-phase) $1b_1$, $3a_1$, and $1b_2$, according to reference[4]. The arrow highlights PE signatures due to $Mg^{2+}$–phosphate interactions, as explained in the text. The Mg 2s, P 2p, and P 2s data are shown with a vertical offset for a better comparison, and chemical shifts are highlighted by the dashed lines. The P 2p and P 2s data were normalized in intensity, considering that the $ATP_{(aq)}$ concentration remains constant across those data sets (as opposed to the $Mg^{2+}_{(aq)}$ concentration in the Mg 2p and Mg 2s spectra, whose variation is reflected in the peak intensities). Linear baselines were subtracted from the P 2s data to remove the secondary electron background (see Figure S5 in the SI for details).



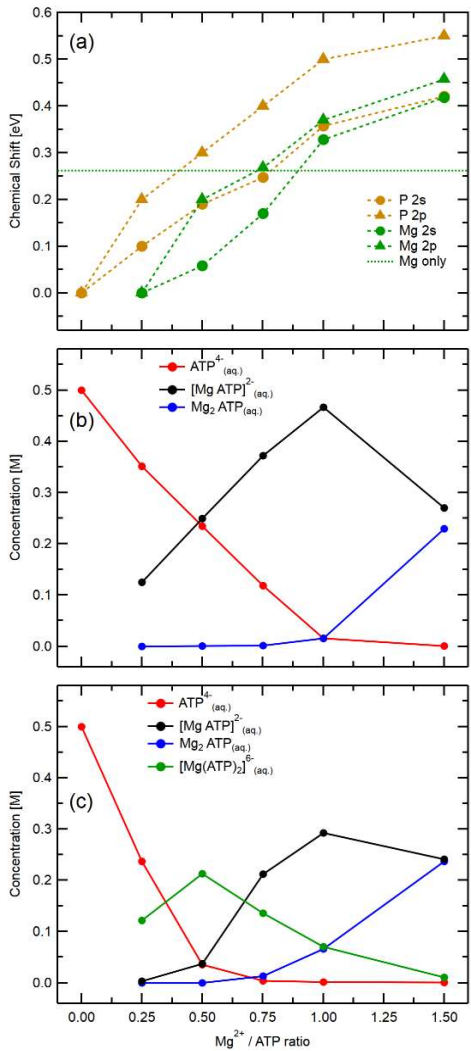

**Figure 5.** (a) Chemical shifts extracted from measurements of the Mg 2p, Mg 2s, P 2p, and P 2s core levels as a function of the $Mg^{2+}$/ATP concentration ratio. A value measured for a 0.5 M $Mg(NO_3)_{2(aq)}$ solution without ATP is shown in green-dotted line for reference. (b) and (c) Molar concentration of the predominant species – $ATP^{4-}_{(aq)}$, $[MgATP]^{2-}_{(aq)}$, $Mg_2ATP_{(aq)}$, and $[Mg(ATP)_2]^{6-}_{(aq)}$ – as a function of the $Mg^{2+}$/ATP concentration ratio at a solution pH of 8.2, calculated within this work using equilibrium constants collected in reference [25] and reference [28]. Details on the two procedures can be found in the SI.

We start the discussion of the core-level spectra with a comment on the solution composition of the $Mg^{2+}$/ATP$_{(aq)}$ samples studied here using the binding constants collected in reference [27] and the calculation of different $Mg^{2+}$–ATP$_{(aq)}$ complexes reported there (see the SI for details, including Table S2). While these values refer to lower ionic strength conditions compared to the samples studied in this work, our primary focus is to produce a first-order approximation speciation plot to guide our qualitative description of the data. Using the consistent data set of equilibrium constants, $ATP^{4-}_{(aq)}$ and $[MgATP]^{2-}_{(aq)}$ are characterized as the dominant species at $Mg^{2+}$/ATP concentration ratios between 0.25:1 and 0.75:1, with an increasing proportion of $[MgATP]^{2-}_{(aq)}$ as the concentration of $Mg^{2+}$ increases. At a 1:1 $Mg^{2+}$/ATP ratio, there is no longer free ATP$_{(aq)}$ ($ATP^{4-}_{(aq)}$) in solution, and $[MgATP]^{2-}_{(aq)}$ is the prevailing species. At a 1.5:1 $Mg^{2+}$/ATP ratio, the $[MgATP]^{2-}_{(aq)}$ concentration decreases as the formation of $[Mg_2ATP]_{(aq)}$ becomes relevant. We remind the reader that the amount of free $Mg^{2+}_{(aq)}$ is negligible at all the $Mg^{2+}$/ATP concentration ratios studied here. The results are summarized in Figure 5(b).

We consider this a minimal set of species expected in significant amounts in the prepared solutions as suggested by the speciation analysis of reference [27]. However, based on subsequently observed $^{25}Mg$ NMR peak broadening for solutions with [ATP] > [$Mg^{2+}$], Bock et al.[28] concluded that an additional species must be taken into account, in the case of similar or higher [ATP]/[$Mg^{2+}$]. Using apparent association constants from that work, we calculated an extended speciation plot that includes $[Mg(ATP)_2]^{6-}_{(aq)}$; see Figure 5(c). The picture is now somewhat different from our initial figure, with the $[Mg(ATP)_2]^{6-}_{(aq)}$ species being dominant for low $Mg^{2+}$ concentrations[23]. We will show that measured chemical shifts can indeed be used to probe speciation, and we can in particular determine which of the two reported models better matches the experiment.

The different species have different manifestations in the PE spectrum. Based on simple electrostatic reasoning, we could expect an electron stabilization (increase of the BE) for the electrons in the P atom upon complexation with $Mg^{2+}$. This is generally confirmed by our P 2s calculations presented in Table 3 in the following section. On the other hand, the electrons in $Mg^{2+}$ should be destabilized by the interaction with the negative charge of ATP$_{(aq)}$. To confirm the expected trends, we performed calculations of Mg 2s and 2p BEs for different species. The results are presented in Table 2. Our computations reveal a decrease in BEs when moving from pure $Mg^{2+}_{(aq)}$ to $[MgATP]^{2-}_{(aq)}$ and an increase in BE when moving to $[Mg_2ATP]_{(aq)}$. The predicted effect on chemical shifts is approximately the same for Mg 2s and Mg 2p.

This is in qualitative, yet to be detailed, agreement with the trend observed for the Mg 2s data in Figure 5(a). Core-level BEs were determined from Voigt fits with a cubic background (Mg 2s, Mg 2p and P 2s), or from reading the positions of the peak maxima (P 2p). For Mg 2s and P, this figure was compiled from short-range scans of the core levels, as detailed in the Methods Section. For the $Mg^{2+}_{(aq)}$-only solution, our calibrated BE (see green-dotted horizontal line) is in good agreement with a measurement of a 3 M $MgCl_{2(aq)}$ solution reported earlier (94.47 versus 94.50 eV),[84] if the liquid water $1b_1$ BE used in that work is set to 11.33 eV, the value used here. Generally, however, we must expect an inherent experimental error in determining accurate absolute BEs when calibrating to peak positions measured from neat liquid water, as we are not accounting for solute-induced effects on the water electronic structure; see again footnote §.



As we detail in the SI the magnitude of the associated spectral shifts is however typically smaller than the variations seen in Figure 5(a). Although uncertainties in the absolute BEs are thus up to 200 meV, the individual error in each data point on the relative scale used in the figure is below 0.1 eV (±0.05 eV).

| Table 2. Calculated $Mg^{2+}_{(aq)}$ and $Mg^{2+}$–$ATP_{(aq)}$ Mg 2s and 2p BEs (in eV) | | |
|---|---|---|
|  | Mg 2s | Mg 2p |
| $Mg^{2+}_{(aq)}$ | 96.01 | 55.24 |
| $[Mg(ATP)_2]^{6-}_{(aq)}$ | 94.03 | 53.27 |
| $[MgATP]^{2-}_{(aq)}$ | 95.21 | 54.42 |
| $[Mg_2ATP]_{(aq)}$ | 96.08 | 55.30 |

We can now test the observed concentration dependence against the aforementioned speciation models, Figures 5(b) and 5(c). Following the Occam's razor principle, we will start with the minimal model, assuming the $[MgATP]^{2-}_{(aq)}$ species is the totally dominating form up to the 1:1 $Mg^{2+}$/ATP concentration ratio (Figure 5(b)). The P core-level BE should then increase monotonically with concentration. This is indeed the case (see Figure 5(a)); the signal exhibits a clear positive chemical shift (*i.e.*, towards higher BEs), up to 500 meV for the highest $Mg^{2+}$/ATP concentration ratios. The Mg core-level BE should drop when moving from $Mg^{2+}$ in water to the complex (data points below the green dashed line), and then it should stay constant. While we see the initial drop, the Mg core-level BEs increase with almost the same slope as the P core-level data.

We conclude that more species are involved in the solutions, consistent with the NMR measurements by Bock et al.[28] Indeed, the extended speciation plot in Figure 5(c) is in qualitative agreement with the observed chemical shifts. As before, the P core-level BE increases for $[MgATP]^{2-}_{(aq)}$ and $[Mg_2ATP]_{(aq)}$ (corresponding to higher $Mg^{2+}$/ATP concentration ratios). However, there is no predicted shift between $ATP^{4-}_{(aq)}$ and the new species $[Mg(ATP)_2]^{6-}_{(aq)}$ (corresponding to lower $Mg^{2+}$/ATP ratios) in Table 3, while it is observed experimentally in Figure 5(a). Nevertheless, such computational discrepancies can be expected as we are comparing the two species with the highest negative charges and the calculations do not include the $Na^+$ counter-ion, affecting the BEs of electrons in the phosphate chain.[6] Since there is a mixture of these species with free $ATP_{(aq)}$, the shift is gradual.

On the other hand, the shifting of the Mg core-level peaks is even more interesting. The $[Mg(ATP)_2]^{6-}_{(aq)}$ species, formed at the lowest concentration, leads to the destabilization of the $Mg^{2+}$ electrons when compared to the $Mg^{2+}$ ion in water. As seen in Table 2, the destabilization is smaller for $[MgATP]^{2-}_{(aq)}$, which is prevalent at higher concentrations, and the BE further grows for $[Mg_2ATP]_{(aq)}$. Our *ab initio* calculations reveal, however, certain quantitative differences compared to the experiment: the BE destabilization of $Mg^{2+}_{(aq)}$ upon complexation with one or two $ATP_{(aq)}$ units is predicted to be larger (800 meV and 2000 meV) than what is observed (350 meV and 440 meV), and the chemical shift for $[Mg_2ATP]_{(aq)}$ is predicted to be too small. In both cases, the $Mg^{2+}_{(aq)}$ electrons are over stabilized which could be attributed to the presence of $Na^+$ ions (not covered in the calculations) and the computational limitations – describing solvation of highly charged anions with the dielectric continuum model and using only the Hartee–Fock electronic structure theory due to the size of the system.

**α-, β-, and γ-phosphate-specific interactions in $ATP_{(aq)}$ and $Mg^{2+}$–$ATP_{(aq)}$.** In the previous section we have inspected global chemical shifts, P 2s and P 2p. We now explore whether we can decompose the P-peak shape to distinguish between contributions between α-, β-, and γ-phosphate.

As a starting point, we performed computations to calculate the P 2s BEs of each phosphate unit in fully deprotonated $ATP_{(aq)}$ ($ATP^{4-}_{(aq)}$), as found at the solution pH of the samples studied here. We obtained nearly identical α- and β-phosphate P 2s BEs (196.86 eV and 196.85 eV, respectively), and an ~880 meV lower γ-phosphate P 2s BE (195.98 eV). In agreement with a previous report on P 2p PES from solid-phase calcium tripolyphosphate,[85] our results show that the bridging phosphate groups (α and β units) have higher BEs than the terminal phosphate (γ-phosphate). In $ATP^{4-}_{(aq)}$, P core-level electrons in the γ-phosphate are uniquely subjected to repulsive interactions from two negatively charged O sites, while α and β units contain a single deprotonated site [see Figure 1(a)]. Such differences in the local chemical environment lead to lower γ-phosphate P 2s BE values compared to the α and β units. (We note that the present and following discussions do not extend to P 2p BEs since the expected doublet peak-structure would prevent a similarly accurate distinction of the small energetic differences associated with the specific phosphate groups.)

Having computationally explored the magnitude of the differences in BE between each phosphate unit in $ATP_{(aq)}$, we attempted to deconvolve P 2s PE spectra recorded from $ATP_{(aq)}$ samples into individual α-, β-, and γ-phosphate contributions with the aid of spectra from $AMP_{(aq)}$ and $ADP_{(aq)}$. When discussing the respective electron BEs, it is useful to refer to each phosphate group as adenosine–phosphate bridging unit, phosphate–phosphate bridging unit and terminal unit, rather than α, β, and γ, respectively; this characterization most directly reflects the origin of the different BEs.

Figure 6 shows P 2s PE spectra from 0.5 M $AMP_{(aq)}$ (magenta curve), $ADP_{(aq)}$ (cyan curve), and $ATP_{(aq)}$ (black curve) solutions at pH 8.2 recorded using a photon energy of 330 eV. A linear baseline was subtracted from each spectrum, and the BE scale was calibrated with respect to the liquid water $1b_1$ feature[3] in valence data recorded under similar conditions. P 2s peak areas extracted from Voigt profile fits were used to re-scale the signal intensities of the $ADP_{(aq)}$ and $ATP_{(aq)}$ spectra to twice ($ADP_{(aq)}$) and three times ($ATP_{(aq)}$) the intensity of the $AMP_{(aq)}$ spectrum, based



on the number of phosphate units in each case. Further details regarding the data treatment are presented in Figure S6 in the SI.

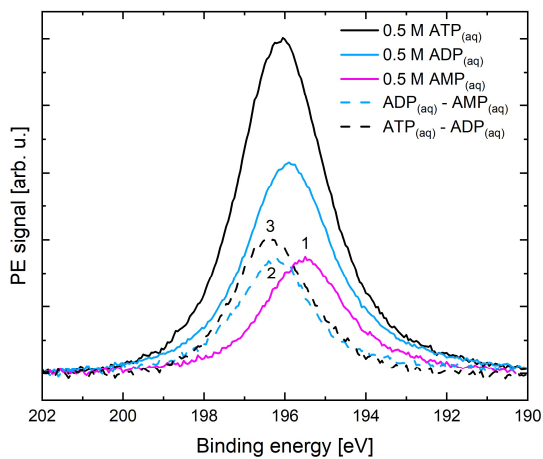

**Figure 6.** P 2s PE spectra recorded from $AMP_{(aq)}$, $ADP_{(aq)}$, and $ATP_{(aq)}$ solutions without dissolved $Mg^{2+}$. The $AMP_{(aq)}$ data is representative of a terminal phosphate (peak 1), while bridging phosphate units in $ATP_{(aq)}$ (peaks 2 and 3) were identified by subtracting the $AMP_{(aq)}$ data from the $ADP_{(aq)}$ spectrum (cyan dashed line) and the $ADP_{(aq)}$ data from the $ATP_{(aq)}$ spectrum (black dashed line).

A BE of 195.3 eV (peak 1 in Figure 6) was determined for $AMP_{(aq)}$, corresponding to its single (terminal) phosphate group. Given that the $ADP_{(aq)}$ phosphate chain contains two phosphate units, a second phosphate BE (peak 2) was determined by subtracting the $AMP_{(aq)}$ spectrum from the $ADP_{(aq)}$ data. The difference spectrum is shown using a cyan dashed line. A similar approach was implemented to determine the BE of a third phosphate unit (peak 3) by producing the difference spectrum, $ATP_{(aq)}$ minus $ADP_{(aq)}$, shown as black dashed line in Figure 5. Voigt profile fits to the difference spectra reveal BEs of 196.2 eV and 196.3 eV for peak 2 and 3, respectively (see Figure S6 in the SI).

As mentioned in the experimental section, we consider the methodology applied here to provide relative BE values rather than absolute values. Hence, we determined BE energy differences of 900 meV between peaks 1 and 2, and 100 meV between peaks 2 and 3, in agreement with our computations. Based on these results, we assign peak 1 as a PE signature of the terminal phosphate (γ) in $ATP_{(aq)}$, and peaks 2 and 3 as PE signatures of bridging phosphate units. In doing so, we are assuming that the terminal phosphate P 2s BE value is the same in $AMP_{(aq)}$, $ADP_{(aq)}$, and $ATP_{(aq)}$. The other two phosphate groups, the adenosine–phosphate bridging (α) moiety and the phosphate–phosphate bridging (β) unit, correspond each to different chemical environments, associated with different P 2s BEs.

To assign the BEs extracted from peaks 2 and 3 to such moieties, and to evaluate our assumption more generally, we calculated the P 2s BEs of the individual phosphate units in $ADP_{(aq)}$ ($[MgADP]^{-}_{(aq)}$). Our results confirm that the terminal phosphate in $ADP_{(aq)}$ has a lower BE (196.04 eV) compared to the bridging phosphate attached to the nucleoside (196.80 eV). With that in mind, we can assign peaks 2 and 3 as β- and α-phosphate in $ATP_{(aq)}$, respectively. Our calculations also show that our assumption of the BE values of the terminal phosphate units in $ADP_{(aq)}$ and $ATP_{(aq)}$ being of the same magnitude is valid.

In addition, we performed calculations to investigate the effect of $Mg^{2+}$–phosphate interactions on the P core-level BEs of each individual phosphate group in $ATP_{(aq)}$ by calculating the P 2s BEs of α-, β-, and γ- phosphate of $[Mg(ATP)_2]^{6-}_{(aq)}$, $[MgATP]^{2-}_{(aq)}$ and $Mg_2ATP_{(aq)}$ (compare Figure 5(b), (c)). For $[MgATP]^{2-}_{(aq)}$, we considered two different binding motifs – the cation simultaneously bound to the three phosphate units and the cation bound only to the β- and γ-phosphate groups. The results are summarized in Table 3. Overall, we observe an increase in the P 2s BEs due to $Mg^{2+}$–phosphate interactions, in agreement with the experimental results from the previous section. For all phosphate units in $[MgATP]^{2-}_{(aq)}$, changes in the number of phosphate groups interacting with $Mg^{2+}_{(aq)}$ lead to BE differences of ~100 meV between the two binding motifs cases. On the other hand, binding to a second $Mg^{2+}_{(aq)}$ ion, as in $Mg_2ATP_{(aq)}$, results in the largest increment in the P 2s BEs, as expected due to the additional positive charge from the metal cation. In addition, the presence of $Mg^{2+}$–phosphate interactions in $[MgATP]^{2-}_{(aq)}$ and in $Mg_2ATP_{(aq)}$ causes the α- and β-phosphate P 2s BEs to adopt different values, as opposed to $ATP_{(aq)}$ in the absence of the divalent cation. Moreover, Table 3 suggests that the energy changes are large enough to be accessed by experiment. But such an attempt would require absolute BEs determination. This will be possible in future studies and opens the possibility of direct determining association equilibrium constants and associated free energies, and thus the effect of divalent cation binding on the ATP hydration free energy, based on careful peak-shape analysis.

**Table 3.** Calculated $ATP_{(aq)}$ and $Mg^{2+}$–$ATP_{(aq)}$ P 2s α-, β-, and γ-phosphate BEs (in eV)

|  | α | b | γ |
| --- | --- | --- | --- |
| $[ATP]^{4-}_{(aq)}$ | 196.86 | 196.85 | 195.98 |
| $[Mg(ATP)_2]^{6-}_{(aq)}$ | 196.98 | 196.78 | 195.94 |
| $[MgATP]^{2-}_{(aq)}$ ($Mg^{2+}$-bonding to α-, β-, and γ-phosphate) | 197.33 | 197.15 | 196.31 |
| $[MgATP]^{2-}_{(aq)}$ ($Mg^{2+}$-bonding to β- and γ-phosphate) | 197.28 | 197.01 | 196.39 |
| $[Mg_2ATP]_{(aq)}$ | 198.27 | 197.83 | 196.82 |

**ICD spectroscopy: Intermolecular-specific probe of the $Mg^{2+}_{(aq)}$ coordination environment in the presence of $ATP_{(aq)}$.** The analysis of the $Mg^{2+}_{(aq)}$ concentration-dependent valence and core-level spectral changes presented



in the previous sections provides site-specific information on the $Mg^{2+}$–$ATP_{(aq)}$ bonding interactions. As explained when introducing Figure 1, ICD spectra, to be presented in the following, are particularly sensitive to interactions between $Mg^{2+}_{(aq)}$ and its immediate coordination environment, revealing additional insight into the number and identity of its chelating units.

We show here that the ICD signal intensity associated with the hydration shell of a charged atomic ion is proportional to the number of hydrating water molecules and, as a result, the ion–water ICD signal decreases upon replacement of a water molecule by a solute component. In this way, we are sensitive to the quantitative exchange of water molecules by $ATP_{(aq)}$ due to the formation of $Mg^{2+}$–$ATP_{(aq)}$; this is illustrated in Figure 7a, where one water molecule is replaced by one phosphate group in the first solvation shell of $Mg^{2+}_{(aq)}$. The associated ICD spectrum further reveals the character of the most involved water orbital,[13] as we discuss later.

The specific ICD process explored here, and illustrated in Figure 7(b), takes place after the initial photoionization of Mg 1s core-level electrons producing primary photoelectrons $e^-_{ph}$ (also compare Figure 1). The Mg 1s core hole left behind is refilled by electrons from the Mg 2s (or Mg 2p) core levels, and the released excess energy is used to ionize the surrounding molecules – $i.e.$, the water molecules and the chelating phosphate units in the first coordination shell – producing ICD electrons $e^-_{ICD}$. (We note that filling the 1s core hole within a local Auger process is more likely, but not of interest here. See reference [13] for a detailed discussion of the core-hole relaxation channels.)

The ICD signal can be well recognized experimentally by constant kinetic energy (KE) of the ICD electrons, $i.e.$, independent of the incident photon energy, just as in the case of Auger electrons. Note further that the ICD signature in the present case occurs in a KE range considerably larger than the respective Auger electrons, which for ICD following $Mg^{2+}_{(aq)}$ 1s ionization is near 1190–1250 eV ($versus$ 1160–1190 eV for the respective KLL Auger peak), with ICD processes involving Mg 2p and Mg 2s near 1238 eV and 1180 eV, respectively.[13] The much higher kinetic energy of the ICD electrons relative to the respective Auger electrons roughly reflects the considerably larger Mg 2p, 2s BEs than the water valence BEs.

Figure 8(a) shows ICD spectra recorded from 0.5 M $Mg(NO_3)_{2(aq)}$ solutions without $ATP_{(aq)}$ (black line) and with 1:1 and 1.5:1 $Mg^{2+}$/ATP concentration ratios (orange and red lines, respectively). For the 0.5 M $Mg(NO_3)_{2(aq)}$ solutions (no ATP added), the cation exists in its octahedral configuration ($[Mg(H_2O)_6]^{2+}_{(aq)}$ (compare Figure 7(a)), regardless of the presence of the counter ion ($NO_3^-_{(aq)}$).[23, 27] The data were recorded at a photon energy of 1314 eV, which was selected to exceed the Mg 1s BE (based on reference[13]). The secondary-electron background (from inelastically scattered (photo)electrons in solution) was removed by fitting and subsequently subtracting cubic baselines from the different spectral regions. The Mg 2s peak areas extracted from Voigt fits to each data set were used to normalize the signal intensity in all the spectra by scaling the PE signal ordinate based on the $Mg^{2+}_{(aq)}$ concentration, assuming equal cross sections. In that way, we normalize the data in Figure 8(a) so as to display the same peak area for the samples where the $Mg^{2+}_{(aq)}$ concentration is 0.5 M (black and orange curves) and an area 1.5 times larger for the sample containing 0.75 M $Mg^{2+}_{(aq)}$ (red curve). The energy scale in Figure 8(a) was calibrated by firstly shifting the $Mg^{2+}_{(aq)}$-only data to an energy of the main K-LL Auger line (not shown in the Figure) of 1175.5 eV ([13], the value for a 2 M aqueous solution of $MgCl_2$). The energy scale for the $Mg^{2+}$/$ATP_{(aq)}$ data was then calibrated such that the Mg 2s kinetic energies are consistent with the BEs shown in Figure 5(a). Further details regarding the data treatment can be found in Figure S7 of the SI.

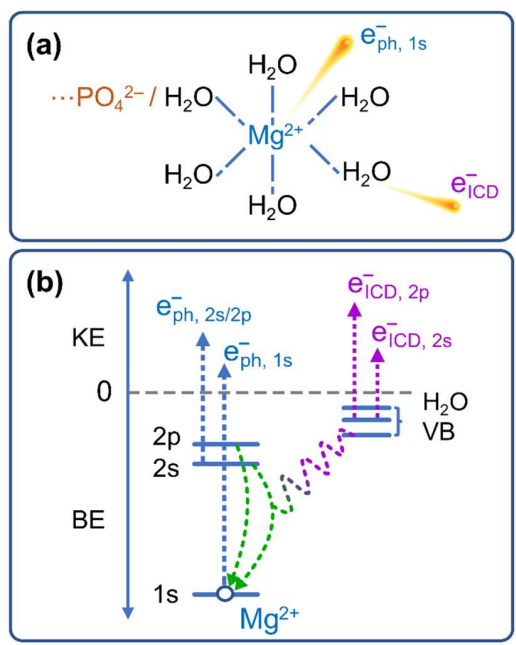

**Figure 7.** (a) Ejection of the initial 1s photoelectron, 1s $e^-_{ph}$, from $Mg^{2+}_{(aq)}$ and an ICD electron, $e^-_{ICD}$, associated with the subsequent valence ionization of a hydration-shell water molecule. The sketch also illustrates the direct interaction of $Mg^{2+}_{(aq)}$ with a phosphate unit of $ATP_{(aq)}$ which reduces the number of water molecules in the first solvation shell. (b) Energy level diagram depicting the ICD process studied here. The open circle denotes the 1s core hole upon ionization of $Mg^{2+}$; $e^-_{ph}$ are the ejected 1s, 2s, 2p photoelectrons. BE and KE denote electron binding and electron kinetic energies of the measured electrons, respectively. The non-local ICD processes, where the relaxation of the metal core hole involves the first solvation shell, lead to the ionization of all water valence (VB) orbitals (and potentially also of phosphate). This produces the ICD electrons $e^-_{ICD,2s}$ and $e^-_{ICD,2p}$.

Our first observation is that the Mg 2p and Mg 2s ICD signals occur at slightly but distinctly different KEs in the $Mg^{2+}$/$ATP_{(aq)}$ spectra with respect to the $Mg^{2+}_{(aq)}$-only



([Mg(H$_2$O)$_6$]$^{2+}_{(aq)}$) data. This directly reflects the formation of [MgATP]$^{2-}_{(aq)}$ (in the 1:1 Mg$^{2+}$/ATP ratio case) and Mg$_2$ATP$_{(aq)}$ (in the 1.5:1 Mg$^{2+}$/ATP ratio case) as discussed in connection with Figure 5(b). The observed spectral changes suggest that interactions with ATP$_{(aq)}$ may also affect the transfer of excess energy from Mg$^{2+}_{(aq)}$ to its coordinating environment during the ICD process, *e.g.,* due to charge redistribution between the metal ion and the phosphate units in the nucleotide. We note that the Mg$^{2+}_{(aq)}$-only sample did not contain Tris$_{(aq)}$, in contrast to the Mg$^{2+}$/ATP solutions, but Mg$^{2+}$–Tris$_{(aq)}$ complexation should be relatively weak, as discussed above.

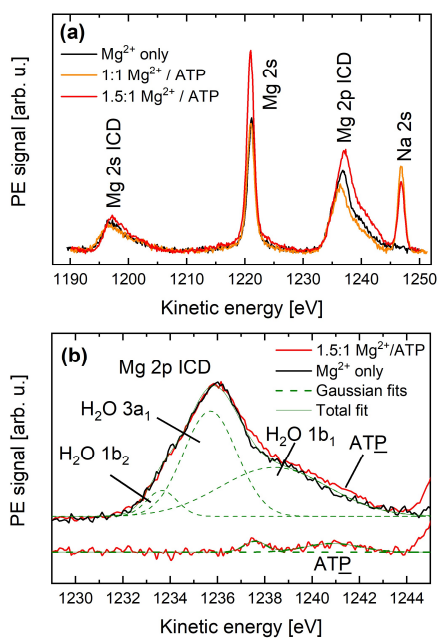

**Figure 8.** (a) ICD spectra from Mg$^{2+}_{(aq)}$ without ATP$_{(aq)}$ and in the presence of ATP$_{(aq)}$ at 1:1 and 1.5:1 Mg$^{2+}$/ATP concentration ratios. The signal intensity has been normalized to the Mg$^{2+}_{(aq)}$ concentration (see text). The spectral range of the 2s and 2p ICD channels covers the direct Mg 2s ionization. (b) Close-up of the Mg 2p ICD spectral region presented in panel (a). Spectra in panel (b) are displayed such that the main peaks are at the same KE, and their heights are the same. Gaussian curves (green dashed lines) highlight the individual ionization features from the water orbitals 1b$_2$, 3a$_1$ and 1b$_1$ involved in the ICD process of the Mg$^{2+}$ only data. The Mg$^{2+}$/ATP minus the Mg$^{2+}$-only difference spectrum shown at the bottom (also in red) reveals the ICD signal associated with the Mg$^{2+}$–phosphate interaction (labelled AT$\underline{P}$, associated Gaussian fit in green).

Our second observation is that the intensity of the Mg 2p ICD feature decreases in the 1:1 Mg$^{2+}$/ATP data compared to the Mg$^{2+}_{(aq)}$-only data, despite both of them being associated with equal Mg$^{2+}_{(aq)}$ concentrations. This reveals the aforementioned formation of [MgATP]$^{2-}_{(aq)}$ and the associated reduction in the number of the Mg$^{2+}_{(aq)}$–water ICD channels available. Accordingly, we observe an ∼15% reduction in signal intensity. This corresponds to one water molecule out of six being replaced in [Mg(H$_2$O)$_6$]$^{2+}_{(aq)}$ upon complexation to ATP$_{(aq)}$ (assuming negligible Mg$^{2+}$–Tris$_{(aq)}$ complexation); compare Figure 6(a). The ICD spectral finger print associated with the phosphate will be detailed next.

As expected, based on the Mg$^{2+}_{(aq)}$ concentration in each sample, the intensity of the Mg 2p ICD signal in the 1.5:1 Mg$^{2+}$/ATP data is ∼1.5 times higher with respect to the 1:1 Mg$^{2+}$/ATP case. However, if we compare the 1.5:1 Mg$^{2+}$/ATP spectra with the Mg$^{2+}_{(aq)}$-only data, we observe that the ICD signal intensity in the 1.5:1 Mg$^{2+}$/ATP case is only 1.3 times higher than in the Mg$^{2+}_{(aq)}$-only case. This shows that the presence of ATP$_{(aq)}$ causes the signal to decrease with respect to the concentration of the metal cation. In other words, and following our argument from the previous paragraph, we observe an ∼15% reduction in signal due to the formation of [MgATP]$^{2-}_{(aq)}$ and Mg$_2$ATP$_{(aq)}$.

Another spectral change becomes apparent when overlapping the main peaks (at a common KE and normalized area) of the Mg 2p ICD spectra from both samples, as shown in Figure 8(b) and detailed in the respective figure caption. We recall that the ICD signal is the convolution of ionization peaks associated with all the orbitals involved in the energy transfer process occurring within the coordination shell. The observed broader Mg 2p ICD feature in the Mg$^{2+}$/ATP spectrum with respect to the Mg$^{2+}_{(aq)}$-only spectrum thus implies that the ICD process in the latter involves additional orbitals, other than the water orbitals.[13] In order to validate this conclusion, we subtracted the scaled Mg$^{2+}_{(aq)}$-only data from the 1.5:1 Mg$^{2+}$/ATP data. The difference spectrum is shown in red at the bottom of Figure 8(b), highlighting an ICD component associated with ATP (labelled AT$\underline{P}$) occurring near 1241 eV KE (see Gaussian fits in green). Note the occurrence of another feature near 1238 eV, the assignment of which is not clear at the moment. Gaussian fits of the individual water orbitals (1b$_2$, 3a$_1$, and 1b$_1$) to the Mg$^{2+}_{(aq)}$-only spectrum are shown in dashed lines and are in semi-quantitative agreement with reference [13]. The AT$\underline{P}$ (phosphate) contribution from ATP$_{(aq)}$ interacting with Mg$^{2+}_{(aq)}$, *i.e.*, due to the formation of species as considered in Figures 5(b), (c), is indeed consistent with the lowest phosphate BE being approximately 2 eV lower than that of water 1b$_1$.[8] As reported previously, Mg$^{2+}_{(aq)}$–water ICD signals show a high specificity for water electrons from the 3a$_1$ orbitals.[13] This is also the case for the data presented here, highlighting the sensitivity of the technique to even orbital spatial orientation.

Overall, the sensitivity of the ICD signal intensity to the charge of the solvated ion involved in the solute–water energy-transfer process allows us to uniquely probe charge distribution changes upon replacing water molecules in the coordination sphere of [Mg(H$_2$O)$_6$]$^{2+}_{(aq)}$ by phosphate from ATP$_{(aq)}$ to form different Mg$^{2+}$–ATP$_{(aq)}$ complexes.



**CONCLUSIONS**

We have demonstrated the application of LJ-PES for gaining insight into molecular structure relevant to biophysics and biochemistry. Irradiation of the aqueous-phase sample with soft X-ray photons provides a complex photoemission spectrum resulting from primary and second-order emission photoelectrons that originate from different subcomponents of the molecule and its interaction with its environment. We obtain a comprehensive view on the molecular interactions present in complex solutions even with multicomponent biomolecular systems.

Specifically, we have probed interactions between both the adenine and phosphate units in ATP$_{(aq)}$ with Mg$^{2+}$$_{(aq)}$, as well as the electronic structure of the phosphate chain in ATP$_{(aq)}$, using LJ-PES combined with theoretical calculations, and have correlated our observations to the formation of [MgATP]$^{2-}$$_{(aq)}$, [Mg(ATP)$_2$]$^{6-}$$_{(aq)}$ and Mg$_2$ATP$_{(aq)}$.

Regarding the photoelectron spectroscopy measurements, valence spectra from solutions with different Mg$^{2+}$/ATP concentration ratios reveal the adenine lowest-ionization peak in ATP$_{(aq)}$ to shift in BE as the Mg$^{2+}$$_{(aq)}$ concentration increases, providing direct evidence of Mg$^{2+}$–adenine and Mg$^{2+}$–phosphate interactions in [MgATP]$^{2-}$$_{(aq)}$, [Mg(ATP)$_2$]$^{6-}$$_{(aq)}$, and Mg$_2$ATP$_{(aq)}$. A more detailed correlation of spectral energy shifts occurring from the various Mg-ATP$_{(aq)}$ species is revealed from the Mg 2p, Mg 2s, P 2p, and P 2s core-level photoelectron spectra of both the divalent cation and ATP$_{(aq)}$. The interaction between Mg$^{2+}$$_{(aq)}$ cations and the phosphate chain of ATP$_{(aq)}$ is directly reflected in the measured chemical shift in photoelectron spectra for both Mg and P core-level electrons. We also performed a combined analysis of P 2s PE spectra from ATP$_{(aq)}$, ADP$_{(aq)}$, and AMP$_{(aq)}$, and computed BEs to isolate spectral fingerprints of α-, β-, and γ-phosphate in ATP$_{(aq)}$. Our results reveal that the BEs of the bridging groups in the phosphate chain are higher than those of the terminal phosphate. We additionally calculated the P 2s phosphate-specific BEs of [MgATP]$^{2-}$$_{(aq)}$, [Mg(ATP)$_2$]$^{6-}$$_{(aq)}$, and Mg$_2$ATP$_{(aq)}$, to characterize the effect of different Mg$^{2+}$$_{(aq)}$-binding motifs. These BEs correlate well with the experimental core-level energy shifts.

The ICD study demonstrates the first application of this technique for solving structural problems, with a sensitivity and access to interaction details not accessible by photoelectron spectroscopy. In fact, photoelectron spectra are only very weakly dependent on the intermolecular interactions.[86] This highlights the enormous potential of ICD spectroscopy, and potentially adds an exciting novel tool to probing complex molecular structure in biochemical-relevant aqueous solutions. Since the ICD signal is completely absent for isolated molecules, it inevitably sensitively reports on the intermolecular interactions. Here, we present ICD spectra from ATP$_{(aq)}$ solutions, containing no Mg$^{2+}$$_{(aq)}$ and with a Mg$^{2+}$/ATP ratio of 1.5, allowing us to probe the interactions between the metal cation and a given coordination environment with exceptional sensitivity. Differences in signal intensity and spectral shape for the two solutions identify the replacement of first-hydration-shell water molecules by phosphate from ATP$_{(aq)}$. This suggests a further potential application of ICD-based electron spectroscopy, namely the quantitative determination of solvation-shell constituents.

While we provide a semi-quantitative spectral assignment of Mg$^{2+}$–ATP$_{(aq)}$ interactions that shows the ability to distinguish between two speciation models, such information cannot be used, at present, to derive thermodynamic information such as association constants. It would be tempting though, in conjunction with application of more recent methods to determine absolute BEs, to accurately decompose the P 2s peak shape into contributions from the α, β, and γ units, and experimentally quantify the associated energy shifts arising from interaction with Mg$^{2+}$ ions. Furthermore, the combined information extracted from the spectral assignments, chemical shifts, and ICD phenomena provides a foundation for future temperature-dependent PES experiments that could potentially recover the Mg$^{2+}$–phosphate-dependent Mg$^{2+}$–ATP$_{(aq)}$ association equilibria energies and entropy data.

In summary, we have explored the capabilities of state-of-the-art LJ-PES, including direct PE emission and non-local relaxation processes upon core-level ionization, in characterizing the electronic-structure interactions between ATP and Mg$^{2+}$ in aqueous solution. Future studies should make use of the structure-sensitivity of ICD to include the respective P- and N-induced relaxation processes. One goal in the context of the present work would be looking for solid experimental evidence of a closed-ring structure, with the adenine–Mg$^{2+}$ interaction potentially showing up in the N 1s ICD spectra.

There are various powerful structural techniques providing analogous structure information, Nuclear Magnetic Resonance being the prime example. LJ-PES has certain advantages, *e.g.*, the surface sensitivity of the technique and capturing the range of instantaneous (rather than time-averaged) structures. However, the full potential of the LJ-PES in biophysics still needs to be explored. It is *e.g.*, imperative to develop empirical guidelines for interpretation of the data without the assistance of *ab initio* theory. It is particularly important to investigate the scope of applicability of the newly emerging ICD spectroscopy, with no previous data to compare. We believe that LJ-PES has a considerable potential to advance complex speciation in biomolecular systems.

**ASSOCIATED CONTENT**

**Supporting Information**. Additional details regarding sample preparation and predicted speciation, computational methods, treatment and analysis of the PE spectra, and the molecular structures and supporting PE data discussed in the text (PDF). This material is available free of charge via the Internet at http://pubs.acs.org. The raw data relevant to this work has been deposited at DOI: 10.5281/zenodo.7998786.

AUTHOR INFORMATION

**Corresponding Authors**

*petr.slavicek@vscht.cz, *stephen.bradforth@usc.edu, *winter@fhi-berlin.mpg.de

**Present Addresses**
†Gas Phase Chemical Physics, Sandia National Laboratories, Livermore, California 94550, USA
*§ Fritz-Haber-Institut der Max-Planck-Gesellschaft, Faradayweg 4-6, 14195 Berlin, Germany*

**Author Contributions**
‡These authors contributed equally.

**Notes**
§ In the present study, no low-energy cut-off measurements from an electrically biased liquid jet have been performed. This is the reason why we cannot apply the more robust method for a determination of absolute electron binding energies, as recently reported in reference[3].

## ACKNOWLEDGMENT
The authors would like to thank Robert Seidel, Christi Schroeder, Anne B. Stephansen, Claudia Kolbeck, Marvin Pohl, Iain Wilkinson, Stephan Thürmer, and Gerard Meijer for their support along different stages of the project and during initial experiments at BESSY II. C.L and D.M.N. were supported by the Director, Office of Basic Energy Science, Chemical Sciences Division of the U.S. Department of Energy under Contract No. DE-AC02-05CH11231 and by the Alexander von Humboldt Foundation. L.T. and P.S. acknowledge the support of the Czech Science Foundation, project number 21-26601X (EXPRO project). S.B. acknowledges NSF CHE-0617060 for early phases of this work and NSF CHE-1665532 for travel support in finalizing the work. B.W. acknowledges funding from the European Research Council (ERC) under the European Union's Horizon 2020 research and innovation programme (grant agreement No. 883759, AQUACHIRAL). F.T. acknowledges funding by the Deutsche Forschungsgemeinschaft (DFG, German Research Foundation) – Project 509471550, Emmy Noether Programme. F.T and B.W. acknowledge support from the MaxWater initiative of the Max-Planck-Gesellschaft. We acknowledge DESY (Hamburg, Germany), a member of the Helmholtz Association HGF, for the provision of experimental facilities. This research was carried out initially at BESSY II (project number 2009_2_90613 at beamline U41-PGM) and subsequently at PETRA III, and we thank Moritz Hoesch as well as the whole beamline staff, the PETRA III chemistry-laboratory and crane operators for assistance in using the P04 soft X-ray beamline.

## ABBREVIATIONS
ADP, adenosine diphosphate; AMP, adenosine monophosphate; ATP, adenosine triphosphate; BE, binding energy; HOMO, highest occupied molecular orbital; HF, Hartree–Fock; ICD, intermolecular Coulombic decay; KE, kinetic energy; LJ-PES, liquid-jet photoelectron spectroscopy; MP2, second-order Møller–Plesset; NMR, nuclear magnetic resonance; PCM, polarizable continuum model; PE, photoelectron; PEEK, polyether ether ketone; PES, photoelectron spectroscopy; Tris, (tris(hydroxymethyl)aminomethane).

## REFERENCES
1. Reinert, F. and S. Hüfner, *Photoemission spectroscopy—from early days to recent applications.* New Journal of Physics, 2005. **7**(1): p. 97.
2. Faubel, M., S. Schlemmer, and J.P. Toennies, *A molecular beam study of the evaporation of water from a liquid jet.* Zeitschrift für Physik D Atoms, Molecules and Clusters, 1988. **10**(2): p. 269-277.
3. Thürmer, S., et al., *Accurate Vertical Ionization Energy and Work Function Determinations of Liquid Water and Aqueous Solutions.* Chemical Science, 2021. **12**(31): p. 10558-10582.
4. Winter, B., et al., *Full Valence Band Photoemission from Liquid Water Using EUV Synchrotron Radiation.* The Journal of Physical Chemistry A, 2004. **108**(14): p. 2625-2632.
5. Mudryk, K.D., et al., *The electronic structure of the aqueous permanganate ion: aqueous-phase energetics and molecular bonding studied using liquid jet photoelectron spectroscopy.* Physical Chemistry Chemical Physics, 2020. **22**(36): p. 20311-20330.
6. Pluhařová, E., et al., *Transforming anion instability into stability: contrasting photoionization of three protonation forms of the phosphate ion upon moving into water.* J Phys Chem B, 2012. **116**(44): p. 13254-64.
7. Malerz, S., et al., *Following in Emil Fischer's Footsteps: A Site-Selective Probe of Glucose Acid–Base Chemistry.* The Journal of Physical Chemistry A, 2021. **125**(32): p. 6881-6892.
8. Schroeder, C.A., et al., *Oxidation Half-Reaction of Aqueous Nucleosides and Nucleotides via Photoelectron Spectroscopy Augmented by ab Initio Calculations.* Journal of the American Chemical Society, 2015. **137**(1): p. 201-209.
9. Dupuy, R., et al., *Core level photoelectron spectroscopy of heterogeneous reactions at liquid–vapor interfaces: Current status, challenges, and prospects.* The Journal of Chemical Physics, 2021. **154**(6): p. 060901.
10. Seidel, R., B. Winter, and S.E. Bradforth, *Valence Electronic Structure of Aqueous Solutions: Insights from Photoelectron Spectroscopy.* Annual Review of Physical Chemistry, 2016. **67**(1): p. 283-305.
11. Pohl, M.N., et al., *Do water's electrons care about electrolytes?* Chemical science, 2019. **10**(3): p. 848-865.
12. Cederbaum, L.S., J. Zobeley, and F. Tarantelli, *Giant Intermolecular Decay and Fragmentation of Clusters.* Physical Review Letters, 1997. **79**(24): p. 4778-4781.
13. Gopakumar, G., et al., *Probing Aqueous Ions with Non-local Auger Relaxation.* Physical Chemistry Chemical Physics, 2022. **24**: p. 8661-8671.
14. Jahnke, T., et al., *Interatomic and Intermolecular Coulombic Decay.* Chemical Reviews, 2020. **120**(20): p. 11295-11369.
15. Langen, P. and F. Hucho, *Karl Lohmann and the Discovery of ATP.* Angewandte Chemie International Edition, 2008. **47**(10): p. 1824-1827.
16. Boyer, P.D., *Energy, life, and ATP. Nobel Lecture, December 8, 1997*, in *Nobel Lectures Chemistry 1996 - 2000*, I. Grenthe, Editor. 2003, World Scientific.
17. Kamerlin, S.C.L., et al., *Why nature really chose phosphate.* Quarterly Reviews of Biophysics, 2013. **46**(1): p. 1-132.
18. Müller, W.E.G., H.C. Schröder, and X. Wang, *Inorganic Polyphosphates As Storage for and Generator of Metabolic Energy in the Extracellular Matrix.* Chemical reviews, 2019. **119**(24): p. 12337-12374.
19. Bonora, M., et al., *ATP synthesis and storage.* Purinergic Signalling, 2012. **8**(3): p. 343-357.
20. Zimmerman, J.J., A. von Saint André-von Arnim, and J. McLaughlin, *Chapter 74 - Cellular Respiration*, in *Pediatric Critical Care (Fourth Edition)*, B.P. Fuhrman and J.J. Zimmerman, Editors. 2011, Mosby: Saint Louis. p. 1058-1072.

21. Sigel, H. and R. Griesser, *Nucleoside 5′-triphosphates: self-association, acid–base, and metal ion-binding properties in solution.* Chemical Society Reviews, 2005. **34**(10): p. 875-900.

22. Tribolet, R. and H. Sigel, *Influence of the protonation degree on the self-association properties of adenosine 5′-triphosphate (ATP).* European Journal of Biochemistry, 1988. **170**(3): p. 617-626.

23. Lightstone, F.C., et al., *A first principles molecular dynamics simulation of the hydrated magnesium ion.* Chemical Physics Letters, 2001. **343**(5): p. 549-555.

24. Glonek, T., *31P NMR of Mg-ATP in dilute solutions: Complexation and exchange.* International Journal of Biochemistry, 1992. **24**(10): p. 1533-1559.

25. Wilson, J.E. and A. Chin, *Chelation of divalent cations by ATP, studied by titration calorimetry.* Analytical Biochemistry, 1991. **193**(1): p. 16-19.

26. Buelens, F.P., et al., *ATP–Magnesium Coordination: Protein Structure-Based Force Field Evaluation and Corrections.* Journal of Chemical Theory and Computation, 2021. **17**(3): p. 1922-1930.

27. Storer, A.C. and A. Cornish-Bowden, *Concentration of MgATP2- and other ions in solution. Calculation of the true concentrations of species present in mixtures of associating ions.* Biochem J, 1976. **159**(1): p. 1-5.

28. Bock, J.L., et al., *25Mg NMR Studies of magnesium binding to erythrocyte constituents.* Journal of Inorganic Biochemistry, 1991. **44**(2): p. 79-87.

29. Molla, G.S., et al., *Mechanistic and kinetics elucidation of Mg2+/ATP molar ratio effect on glycerol kinase.* Molecular Catalysis, 2018. **445**: p. 36-42.

30. Szabó, Z., *Multinuclear NMR studies of the interaction of metal ions with adenine-nucleotides.* Coordination chemistry reviews, 2008. **252**(21-22): p. 2362-2380.

31. Bishop, E.O., et al., *A P-31-Nmr Study of Monomagnesium and Dimagnesium Complexes of Adenosine 5'-Triphosphate and Model Systems.* Biochimica Et Biophysica Acta, 1981. **635**(1): p. 63-72.

32. Sari, J.C., et al., *Microcalorimetric study of magnesium-adenosine triphosphate ternary complex.* Journal of bioenergetics and biomembranes, 1982. **14**(3): p. 171-179.

33. SIGEL, H., *Isomeric equilibria in complexes of adenosine 5′-triphosphate with divalent metal ions.* European Journal of Biochemistry, 1987. **165**(1): p. 65-72.

34. Sigel, H. and B. Song, *Solution structures of nucleotide-metal ion complexes. Isomeric equilibria.* Metal ions in biological systems, 1996. **32**: p. 135-135.

35. Frańska, M., et al., *Gas-Phase Internal Ribose Residue Loss from Mg-ATP and Mg-ADP Complexes: Experimental and Theoretical Evidence for Phosphate-Mg-Adenine Interaction.* Journal of the American Society for Mass Spectrometry, 2022. **33**(8): p. 1474-1479.

36. Matthies, M. and G. Zundel, *Hydration and self-association of adenosine triphosphate, adenosine diphosphate, and their 1:1 complexes with magnesium(II) at various pH values: infrared investigations.* Journal of the Chemical Society, Perkin Transactions 2, 1977(14): p. 1824-1830.

37. Manchester, K.L., *Free energy ATP hydrolysis and phosphorylation potential.* Biochemical Education, 1980. **8**(3): p. 70-72.

38. Achbergerová, L. and J. Nahálka, *Polyphosphate - an ancient energy source and active metabolic regulator.* Microbial Cell Factories, 2011. **10**(1): p. 63.

39. Harrison, C.B. and K. Schulten, *Quantum and Classical Dynamics Simulations of ATP Hydrolysis in Solution.* Journal of Chemical Theory and Computation, 2012. **8**(7): p. 2328-2335.

40. Akola, J. and R.O. Jones, *ATP Hydrolysis in Water – A Density Functional Study.* The Journal of Physical Chemistry B, 2003. **107**(42): p. 11774-11783.

41. Wang, C., W. Huang, and J.-L. Liao, *QM/MM Investigation of ATP Hydrolysis in Aqueous Solution.* The Journal of Physical Chemistry B, 2015. **119**(9): p. 3720-3726.

42. Kamerlin, S. and A. Warshel, *On the energetics of ATP hydrolysis in solution.* The journal of physical chemistry. B, 2009. **113 47**: p. 15692-8.

43. Weber, J. and A.E. Senior, *ATP synthase: what we know about ATP hydrolysis and what we do not know about ATP synthesis.* Biochimica et Biophysica Acta (BBA) - Bioenergetics, 2000. **1458**(2): p. 300-309.

44. Liao, J.C., et al., *The conformational states of Mg.ATP in water.* Eur Biophys J, 2004. **33**(1): p. 29-37.

45. Huang, S.L. and M.D. Tsai, *Does the magnesium (II) ion interact with the. alpha.-phosphate of ATP? An investigation by oxygen-17 nuclear magnetic resonance.* Biochemistry, 1982. **21**(5): p. 951-959.

46. Mildvan, A.S., *Role of magnesium and other divalent cations in ATP-utilizing enzymes.* Magnesium, 1987. **6**(1): p. 28-33.

47. Rajendran, T.E. and T. Muthukumarasamy, *Thermodynamic calculations of biochemical reaction systems at specified pH, pMg, and change in binding of hydrogen and magnesium ions.* Asia-Pacific Journal of Chemical Engineering, 2018. **13**(4): p. e2205.

48. Starikov, E.B., I. Panas, and B. Nordén, *Chemical-to-Mechanical Energy Conversion in Biomacromolecular Machines: A Plasmon and Optimum Control Theory for Directional Work. 1. General Considerations.* The Journal of Physical Chemistry B, 2008. **112**(28): p. 8319-8329.

49. George, P., et al., *"Squiggle-H2O". An enquiry into the importance of solvation effects in phosphate ester and anhydride reactions.* Biochimica et Biophysica Acta (BBA) - Bioenergetics, 1970. **223**(1): p. 1-15.

50. Shikama, K., *Standard free energy maps for the hydrolysis of ATP as a function of pH, pMg and pCa.* Archives of Biochemistry and Biophysics, 1971. **147**(1): p. 311-317.

51. Holm, N., *The significance of Mg in prebiotic geochemistry.* Geobiology, 2012. **10**: p. 269-79.

52. Admiraal, S.J. and D. Herschlag, *Mapping the transition state for ATP hydrolysis: implications for enzymatic catalysis.* Chemistry & Biology, 1995. **2**(11): p. 729-739.

53. Alberty, R.A., *Enzymes: Units of biological structure and function.* Journal of Chemical Education, 1957. **34**(1): p. A33.

54. Sigel, H., *Interactions of metal ions with nucleotides and nucleic acids and their constituents.* Chemical Society Reviews, 1993. **22**(4): p. 255-267.

55. Winter, B. and M. Faubel, *Photoemission from Liquid Aqueous Solutions.* Chemical Reviews, 2006. **106**(4): p. 1176-1211.

56. Ottosson, N., et al., *On the Origins of Core–Electron Chemical Shifts of Small Biomolecules in Aqueous Solution: Insights from Photoemission and ab Initio Calculations of Glycineaq.* Journal of the American Chemical Society, 2011. **133**(9): p. 3120-3130.

57. Nolting, D., et al., *pH-Induced Protonation of Lysine in Aqueous Solution Causes Chemical Shifts in X-ray Photoelectron Spectroscopy.* Journal of the American Chemical Society, 2007. **129**(45): p. 14068-14073.

58. Bruce, J.P. and J.C. Hemminger, *Characterization of Fe2+ Aqueous Solutions with Liquid Jet X-ray Photoelectron Spectroscopy: Chloride Depletion at the Liquid/Vapor Interface Due to Complexation with Fe2+.* The Journal of Physical Chemistry B, 2019. **123**(39): p. 8285-8290.

59. Aziz, E.F., et al., *Interaction between liquid water and hydroxide revealed by core-hole de-excitation.* Nature, 2008. **455**(7209): p. 89-91.

60. Skitnevskaya, A.D., et al., *Two-Sided Impact of Water on the Relaxation of Inner-Valence Vacancies of Biologically Relevant Molecules.* The Journal of Physical Chemistry Letters, 2023: p. 1418-1426.

61. Mathe, Z., et al., *Phosphorus Kβ X-ray emission spectroscopy detects non-covalent interactions of phosphate biomolecules in situ.* Chemical Science, 2021. **12**(22): p. 7888-7901.

62. Lanir, A. and N.T. Yu, *A Raman spectroscopic study of the interaction of divalent metal ions with adenine moiety of adenosine 5′-triphosphate.* Journal of Biological Chemistry, 1979. **254**(13): p. 5882-5887.

63. Cohn, M. and T.R. Hughes, *Nuclear Magnetic Resonance Spectra of Adenosine Di- and Triphosphate: II. EFFECT OF COMPLEXING WITH DIVALENT METAL IONS.* Journal of Biological Chemistry, 1962. **237**(1): p. 176-181.

64. McFadden, R.M.L., et al., *Magnesium(II)-ATP Complexes in 1-Ethyl-3-Methylimidazolium Acetate Solutions Characterized by 31Mg β-Radiation-Detected NMR Spectroscopy.* Angewandte Chemie International Edition, 2022. **61**(35): p. e202207137.

65. Castellani, M.E., D. Avagliano, and J.R.R. Verlet, *Ultrafast Dynamics of the Isolated Adenosine-5′-triphosphate Dianion Probed by Time-Resolved Photoelectron Imaging.* The Journal of Physical Chemistry A, 2021. **125**(17): p. 3646-3652.

66. Shimada, H., et al., *Structural changes of nucleic acid base in aqueous solution as observed in X-ray absorption near edge structure (XANES).* Chemical Physics Letters, 2014. **591**: p. 137-141.

67. Kelly, D.N., et al., *Communication: Near edge x-ray absorption fine structure spectroscopy of aqueous adenosine triphosphate at the carbon and nitrogen K-edges.* The Journal of Chemical Physics, 2010. **133**(10): p. 101103.

68. Wang, P., et al., *Thermodynamic parameters for the interaction of adenosine 5′-diphosphate, and adenosine 5′-triphosphate with Mg2+ from 323.15 to 398.15 K.* Journal of Solution Chemistry, 1995. **24**(10): p. 989-1012.

69. Viefhaus, J., et al., *The Variable Polarization XUV Beamline P04 at PETRA III: Optics, mechanics and their performance.* Nuclear Instruments and Methods in Physics Research Section A: Accelerators, Spectrometers, Detectors and Associated Equipment, 2013. **710**: p. 151-154.

70. Malerz, S., et al., *A setup for studies of photoelectron circular dichroism from chiral molecules in aqueous solution.* Review of Scientific Instruments, 2022. **93**(1): p. 015101.

71. Winter, B., *Liquid microjet for photoelectron spectroscopy.* Nuclear Instruments and Methods in Physics Research Section A: Accelerators, Spectrometers, Detectors and Associated Equipment, 2009. **601**(1): p. 139-150.

72. Gilbert, A.T.B., N.A. Besley, and P.M.W. Gill, *Self-Consistent Field Calculations of Excited States Using the Maximum Overlap Method (MOM).* The Journal of Physical Chemistry A, 2008. **112**(50): p. 13164-13171.

73. Epifanovsky, E., et al., *Software for the frontiers of quantum chemistry: An overview of developments in the Q-Chem 5 package.* The Journal of Chemical Physics, 2021. **155**(8): p. 084801.

74. Macetti, G. and A. Genoni, *Initial Maximum Overlap Method for Large Systems by the Quantum Mechanics/Extremely Localized Molecular Orbital Embedding Technique.* Journal of Chemical Theory and Computation, 2021. **17**(7): p. 4169-4182.

75. Mennucci, B. and J. Tomasi, *Continuum solvation models: A new approach to the problem of solute's charge distribution and cavity boundaries.* The Journal of Chemical Physics, 1997. **106**(12): p. 5151-5158.

76. Cancès, E., B. Mennucci, and J. Tomasi, *A new integral equation formalism for the polarizable continuum model: Theoretical background and applications to isotropic and anisotropic dielectrics.* The Journal of Chemical Physics, 1997. **107**(8): p. 3032-3041.

77. Xu, L. and M.L. Coote, *Improving the Accuracy of PCM-UAHF and PCM-UAKS Calculations Using Optimized Electrostatic Scaling Factors.* Journal of Chemical Theory and Computation, 2019. **15**(12): p. 6958-6967.

78. M. J. Frisch, G.W.T., H. B. Schlegel, G. E. Scuseria, M. A. Robb, J. R. Cheeseman, G. Scalmani, V. Barone, B. Mennucci, G. A. Petersson, H. Nakatsuji, M. Caricato, X. Li, H. P. Hratchian, A. F. Izmaylov, J. Bloino, G. Zheng, J. L. Sonnenberg, M. Hada, M. Ehara, K. Toyota, R. Fukuda, J. Hasegawa, M. Ishida, T. Nakajima, Y. Honda, O. Kitao, H. Nakai, T. Vreven, J. A. Montgomery, Jr., J. E. Peralta, F. Ogliaro, M. Bearpark, J. J. Heyd, E. Brothers, K. N. Kudin, V. N. Staroverov, T. Keith, R. Kobayashi, J. Normand, K. Raghavachari, A. Rendell, J. C. Burant, S. S. Iyengar, J. Tomasi, M. Cossi, N. Rega, J. M. Millam, M. Klene, J. E. Knox, J. B. Cross, V. Bakken, C. Adamo, J. Jaramillo, R. Gomperts, R. E. Stratmann, O. Yazyev, A. J. Austin, R. Cammi, C. Pomelli, J. W. Ochterski, R. L. Martin, K. Morokuma, V. G. Zakrzewski, G. A. Voth, P. Salvador, J. J. Dannenberg, S. Dapprich, A. D. Daniels, O. Farkas, J. B. Foresman, J. V. Ortiz, J. Cioslowski, and D. J. Fox, *Gaussian 09, Revision D.01*. 2013, Gaussian Inc.: Wallingford CT.

79. Yanai, T., D.P. Tew, and N.C. Handy, *A new hybrid exchange–correlation functional using the Coulomb-attenuating method (CAM-B3LYP).* Chemical Physics Letters, 2004. **393**(1): p. 51-57.

80. Winter, B. and M. Faubel, *Photoemission from liquid aqueous solutions.* Chem Rev, 2006. **106**(4): p. 1176-211.

81. Pugini, M., et al., *How to measure work functions from aqueous solutions.* Chemical Science, 2023. **14**(35): p. 9574-9588.

82. Trofimov, A.B., et al., *Photoelectron spectra of the nucleobases cytosine, thymine and adenine.* Journal of Physics B: Atomic, Molecular and Optical Physics, 2006. **39**(2): p. 305.

83. Sherwood, P.M.A., *Introduction to Studies of Phosphorus-Oxygen Compounds by XPS.* Surface Science Spectra, 2002. **9**(1): p. 62-66.

84. Öhrwall, G., et al., *Charge Dependence of Solvent-Mediated Intermolecular Coster−Kronig Decay Dynamics of Aqueous Ions.* The Journal of Physical Chemistry B, 2010. **114**(51): p. 17057-17061.

85. Rizkalla, E.N., M.S. Antonious, and S.S. Anis, *X-ray photoelectron and potentiometric studies of some calcium complexes.* Inorganica Chimica Acta, 1985. **96**(2): p. 171-178.

86. Kumar, G., et al., *The influence of aqueous solvent on the electronic structure and non-adiabatic dynamics of indole explored by liquid-jet photoelectron spectroscopy.* Faraday Discussions, 2018. **212**(0): p. 359-381.

TOC

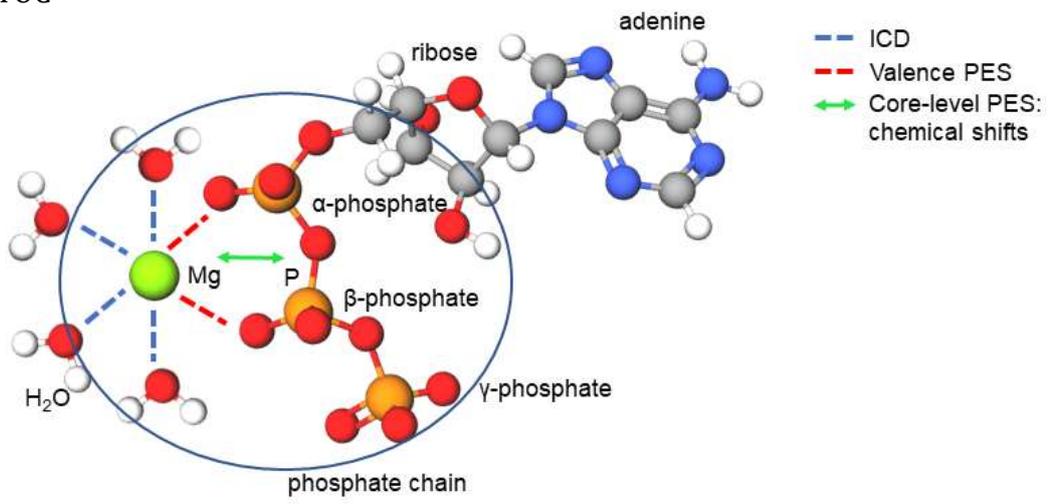

# Supporting Information

# How does $Mg^{2+}_{(aq)}$ interact with $ATP_{(aq)}$? Biomolecular structure through the lens of liquid-jet photoemission spectroscopy

Karen Mudryk[1,#], Chin Lee[1,2,3,a,#], Lukáš Tomaník[4,#], Sebastian Malerz[1], Florian Trinter[1,5], Uwe Hergenhahn[1], Petr Slavíček [4, *] Stephen Bradforth[6,*] & Bernd Winter[1,*]

[1]*Fritz-Haber-Institut der Max-Planck-Gesellschaft, Faradayweg 4-6, 14195 Berlin, Germany*
[2]*Department of Chemistry, University of California, Berkeley, CA 94720, USA*
[3]*Chemical Sciences Division, Lawrence Berkeley National Laboratory, Berkeley, CA 94720, USA*
[4]*Department of Physical Chemistry, University of Chemistry and Technology, Technická 5, Prague 6 16628, Czech Republic*
[5]*Institut für Kernphysik, Goethe-Universität, Max-von-Laue-Straße 1, 60438 Frankfurt am Main, Germany*
[6]*Department of Chemistry, University of Southern California, Los Angeles, CA 90089, USA*

*Current addresses:*
[a] *Gas Phase Chemical Physics, Sandia National Laboratories, Livermore, California 94550, United States*
[#]*Authors contributed equally*
*Corresponding authors: petr.slavicek@vscht.cz, stephen.bradforth@usc.edu, winter@fhi-berlin.mpg.de*

**Table of contents**

## Sample preparation

**Table S1.** Mg(NO$_3$)$_{2(aq.)}$ and Tris$_{(aq)}$ (tris(hydroxymethyl)aminomethane) concentrations in the ATP$_{(aq)}$, ADP$_{(aq)}$, and AMP$_{(aq)}$ solutions used in this work. The solution pH was 8.2 in all samples. All values are in mol dm$^{-3}$.

| Mg$^{2+}$ / ATP ratio | Mg(NO$_3$)$_{2(aq)}$ | Tris$_{(aq)}$ | | |
|---|---|---|---|---|
| | | 0.5 M ATP$_{(aq)}$ | 0.5 M ADP$_{(aq)}$ | 0.5 M AMP$_{(aq)}$ |
| 0 | 0 | 1.16 | 1.36 | 0.23 |
| 0.25 | 0.125 | 1.21 | 1.52 | |
| 0.5 | 0.25 | 1.32 | 1.70 | |
| 0.75 | 0.375 | 1.43 | 1.77 | |
| 1 | 0.5 | 1.54 | 1.93 | |
| 1.5 | 0.75 | 1.65 | 2.34 | |

## Possible 'open form' and 'closed form' configuration of Mg$_2$ATP$_{(aq)}$

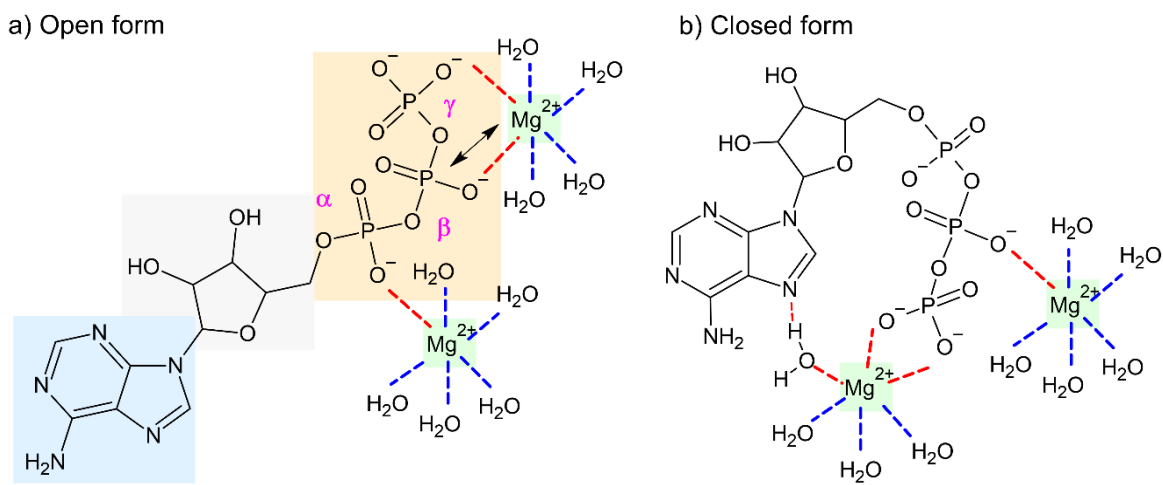

**Figure S1.** Possible molecular structures adopted by Mg$_2$ATP$_{(aq.)}$ a) Open form. b) Close form. This figure was produced using the ACD / ChemSketch software [1].

## Predicted composition of the ATP$_{(aq)}$ samples with dissolved Mg$^{2+}$ as a function of the Mg$^{2+}$/ATP concentration ratio

The composition of the ATP$_{(aq)}$ samples with dissolved Mg$^{2+}$ was calculated as follows. First, we assumed the species in solution to be ATP$^{4-}$, HATP$^{3-}$, H$_2$ATP$^{2-}$, MgATP$^{2-}$, Mg$_2$ATP, MgHATP$^-$, MgH$_2$ATP, Mg$^{2+}$, Na$^+$, H$^+$ and OH$^-$, with a solution pH of 8.2 to match the experimental conditions. Thus,

$$[H^+] = 10^{-8.2} \quad (1)$$

We used equilibrium constants from Storer and Cornish-Bowden [2] – namely, Equations (1), (2), (4), (5), (6) and (7) from Table 2. Notably, the values correspond to solutions of lower ionic strength (0.1 M) than those studied here. Nevertheless, they still serve to allow a qualitative estimate for our system.

$$\frac{[H_2ATP^{2-}]}{[H^+] \cdot [HATP^{3-}]} = 8.5 \cdot 10^3 \quad (2)$$

$$\frac{[\text{HATP}^{3-}]}{[\text{H}^+] \cdot [\text{ATP}^{4-}]} = 1.09 \cdot 10^7 \tag{3}$$

$$\frac{[\text{MgH}_2\text{ATP}]}{[\text{Mg}^{2+}] \cdot [\text{H}_2\text{ATP}^{2-}]} = 2.0 \cdot 10^1 \tag{4}$$

$$\frac{[\text{MgHATP}^-]}{[\text{Mg}^{2+}] \cdot [\text{HATP}^{3-}]} = 5.42 \cdot 10^2 \tag{5}$$

$$\frac{[\text{MgATP}^{2-}]}{[\text{Mg}^{2+}] \cdot [\text{ATP}^{4-}]} = 3.48 \cdot 10^4 \tag{6}$$

$$\frac{[\text{Mg}_2\text{ATP}]}{[\text{Mg}^{2+}] \cdot [\text{MgATP}^{2-}]} = 4.0 \cdot 10^1 \tag{7}$$

In addition, we used equations for mass balance, where $c_0(\text{Na}_2\text{H}_2\text{ATP})$ is the initial concentration of ATP compound and $c_0(\text{Mg(NO}_3)_2)$ is the initial concentration of Mg salt.

$$\begin{aligned} c_0(\text{Na}_2\text{H}_2\text{ATP}) = &[\text{ATP}^{4-}] + [\text{HATP}^{3-}] + [\text{H}_2\text{ATP}^{2-}] + [\text{MgATP}^{2-}] + [\text{Mg}_2\text{ATP}] \\ &+ [\text{MgHATP}^-] + [\text{MgH}_2\text{ATP}] \end{aligned} \tag{8}$$

$$\begin{aligned} c_0(\text{Mg(NO}_3)_2) = &[\text{Mg}^{2+}] + [\text{MgATP}^{2-}] + 2 \cdot [\text{Mg}_2\text{ATP}] + [\text{MgHATP}^-] \\ &+ [\text{MgH}_2\text{ATP}] \end{aligned} \tag{9}$$

$$[\text{Na}^+] = 2 \cdot c_0(\text{Na}_2\text{H}_2\text{ATP}) \tag{10}$$

Finally, we considered the self-ionization of water at 25 °C

$$[\text{H}^+] \cdot [\text{OH}^-] = 10^{-14} \tag{11}$$

This represents a system of 11 equations with 11 unknowns, which was solved for different $Mg^{2+}$/ATP concentration ratios using the Maple 2016 software [3]. The results are reported in Table S2.

The extended speciation plot was created considering the additional complex $Mg(ATP)_2^{6-}$ and the respective equilibrium constant

$$\frac{[\text{MgATP}^{2-}] \cdot [\text{ATP}^{4-}]}{[\text{Mg(ATP)}_2^{6-}]} = 6.2 \cdot 10^{-3} \tag{12}$$

taken from Bock et al.[4] The system of 12 equations with 12 unknowns was solved and the results are reported in Table S3.

**Table S2.** Calculated concentration of species as a function of $Mg^{2+}$/ATP concentration ratio at a solution pH of 8.2. The values were determined as described in the text, using equilibrium constants from Reference [2]. The total concentration of $ATP_{(aq)}$ was kept at 0.5 mol dm$^{-3}$ while the total concentration of $Mg^{2+}_{(aq)}$ was varied in the range 0.125–0.75 mol dm$^{-3}$. All values are in mol dm$^{-3}$.

|  | 0.25:1 ratio | 0.5:1 ratio | 0.75:1 ratio | 1:1 ratio | 1.5:1 ratio |
|---|---|---|---|---|---|
| $ATP^{4-}_{(aq)}$ | 3.51E-01 | 2.34E-01 | 1.18E-01 | 1.57E-02 | 3.66E-04 |
| $H^+_{(aq)}$ | 6.30957E-09 | 6.30957E-09 | 6.30957E-09 | 6.30957E-09 | 6.30957E-09 |
| $[H_2ATP]^{2-}_{(aq)}$ | 1.29E-06 | 8.64E-07 | 4.36E-07 | 5.80E-08 | 1.35E-09 |
| $[HATP]^{3-}_{(aq)}$ | 2.41E-02 | 1.61E-02 | 8.14E-03 | 1.08E-03 | 2.52E-05 |
| $Mg^{2+}_{(aq)}$ | 1.02E-05 | 3.06E-05 | 9.03E-05 | 8.54E-04 | 2.12E-02 |
| $Mg_2ATP_{(aq)}$ | 5.10E-05 | 3.04E-04 | 1.34E-03 | 1.59E-02 | 2.29E-01 |
| $[MgATP]^{2-}_{(aq)}$ | 0.1247542 | 0.249093647 | 0.371824746 | 0.466768697 | 0.270137983 |
| $MgH_2ATP_{(aq)}$ | 2.64455E-10 | 5.2803E-10 | 7.88196E-10 | 9.89459E-10 | 5.7264E-10 |
| $[MgHATP]^-_{(aq)}$ | 1.34E-04 | 2.67E-04 | 3.98E-04 | 5.00E-04 | 2.89E-04 |
| $OH^-_{(aq)}$ | 1.58E-06 | 1.58E-06 | 1.58E-06 | 1.58E-06 | 1.58E-06 |

**Table S3.** Calculated concentration of species (including $Mg(ATP)_2^{6-}$) as a function of $Mg^{2+}$/ATP concentration ratio at a solution pH of 8.2. The values were determined as described in the text, using equilibrium constants from References [2] and [4]. The total concentration of $ATP_{(aq.)}$ was kept at 0.5 mol dm$^{-3}$ while the total concentration of $Mg^{2+}_{(aq)}$ was varied in the range 0.125–0.75 mol dm$^{-3}$. All values are in mol dm$^{-3}$.

|  | 0.25:1 ratio | 0.5:1 ratio | 0.75:1 ratio | 1:1 ratio | 1.5:1 ratio |
|---|---|---|---|---|---|
| $ATP^{4-}_{(aq)}$ | 0.2344094666 | 0.014865245 | 0.00099912 | 0.000509229 | 0.000165 |
| $H^+_{(aq)}$ | 6.30957E-09 | 6.30957E-09 | 6.30957E-09 | 6.30957E-09 | 6.30957E-09 |
| $[H_2ATP]^{2-}_{(aq)}$ | 8.65E-7 | 5.48E-8 | 3.69E-9 | 1.88E-9 | 6.09E-10 |
| $[HATP]^{3-}_{(aq)}$ | 0.01612136 | 0.00102235 | 0.000068714 | 0.00003502 | 0.00001135 |
| $Mg^{2+}_{(aq)}$ | 6.51E-8 | 0.0000304504 | 0.00451273783 | 0.01138740914 | 0.03288926626 |
| $Mg_2ATP_{(aq.)}$ | 1.38E-9 | 0.0000191866 | 0.02832293427 | 0.09191835433 | 0.2484479051 |
| $[MgATP]^{2-}_{(aq)}$ | 0.000530987 | 0.01575231338 | 0.1569054937 | 0.2017982168 | 0.1888518150 |
| $MgH_2ATP_{(aq)}$ | 1.13E-12 | 3.34E-11 | 3.33E-10 | 4.28E-10 | 4.00E-10 |
| $[MgHATP]_{(aq)}$ | 5.69E-7 | 0.0000168729 | 0.0001680679 | 0.000216154 | 0.000202287 |
| $OH^-_{(aq)}$ | 1.58E-06 | 1.58E-06 | 1.58E-06 | 1.58E-06 | 1.58E-06 |
| $[Mg(ATP)_2]^{6-}_{(aq)}$ | 0.1244683764 | 0.2341619901 | 0.1567678318 | 0.1027615107 | 0.03116082140 |

**Photoelectron spectra from Tris$_{(aq)}$ solutions with and without dissolved Mg$^{2+}$**

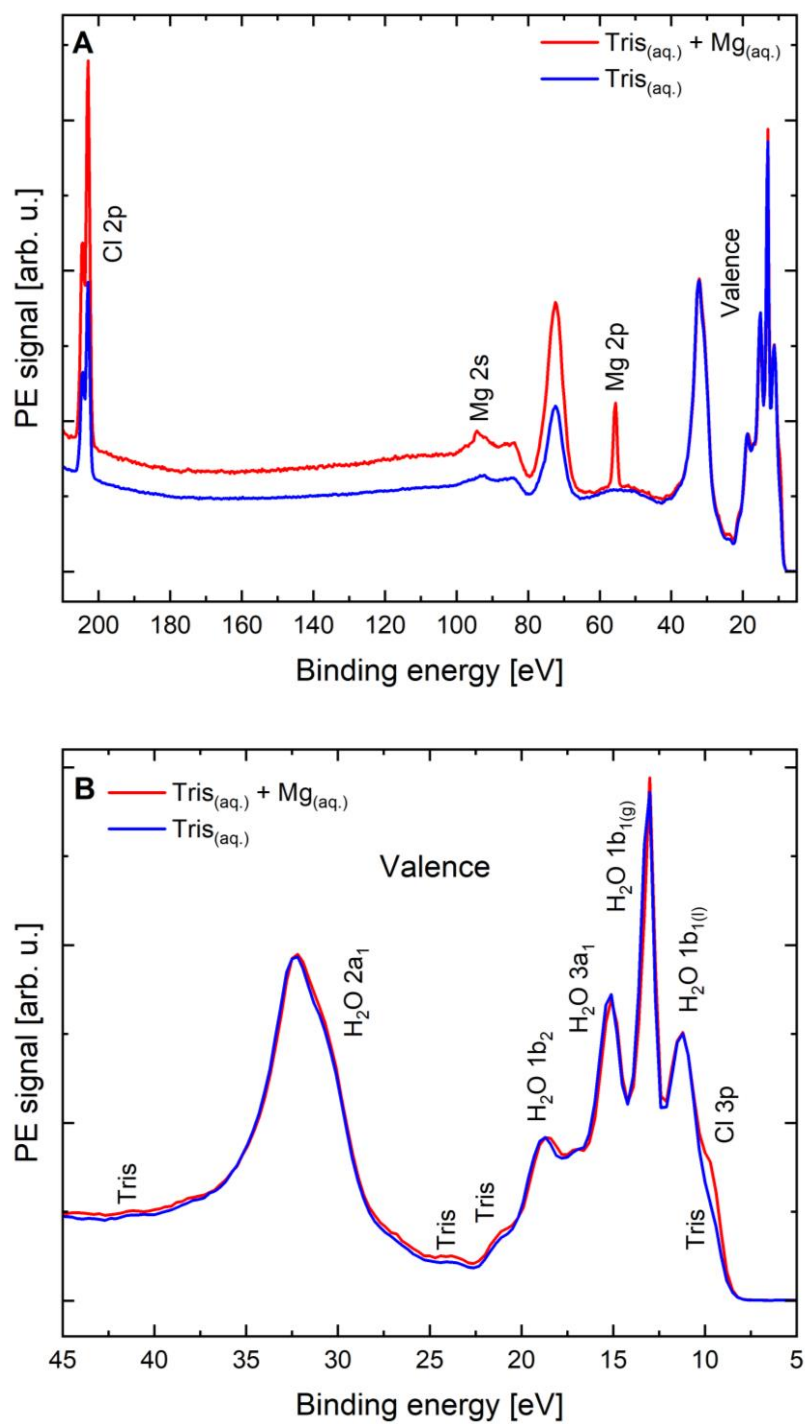

**Figure S2.** a) Photoelectron spectra recorded from 3.2 M Tris$_{(aq)}$ without Mg$^{2+}_{(aq)}$ (blue line) and with 0.5 M MgCl$_{2(aq)}$ (red line) using a photon energy of 250 eV. The sample pH was adjusted to 8.2 by addition of concentrated HCl in both cases. The binding energy (BE) scale was calibrated based on the liquid water solvent valence 1b$_1$ peak position, and spectral intensities are displayed to yield its same height. b) Highlight of the valence spectral region from panel a. Photoelectron signatures of water and Cl$^-_{(aq)}$ are labelled according to References [5] and [6], respectively.

## Valence photoelectron spectra of ATP$_{(aq)}$, ADP$_{(aq)}$ and AMP$_{(aq)}$

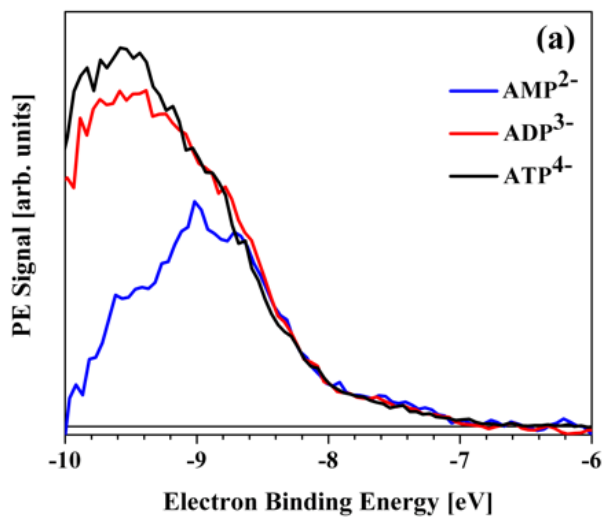

**Figure S3.** Valence photoelectron signals from AMP$_{(aq)}$, ADP$_{(aq)}$, and ATP$_{(aq)}$ solutions without disolved Mg$^{2+}$ isolated by subtracting the solvent water contributions, facilitating the identification of the phosphate feature and its changes in signal intensity across the nucleotide series. The figure was extracted from Reference [7]. The experimental conditions used are described in Reference [8].

## Valence photoelectron spectra of ADP$_{(aq)}$ with added Mg$^{2+}$

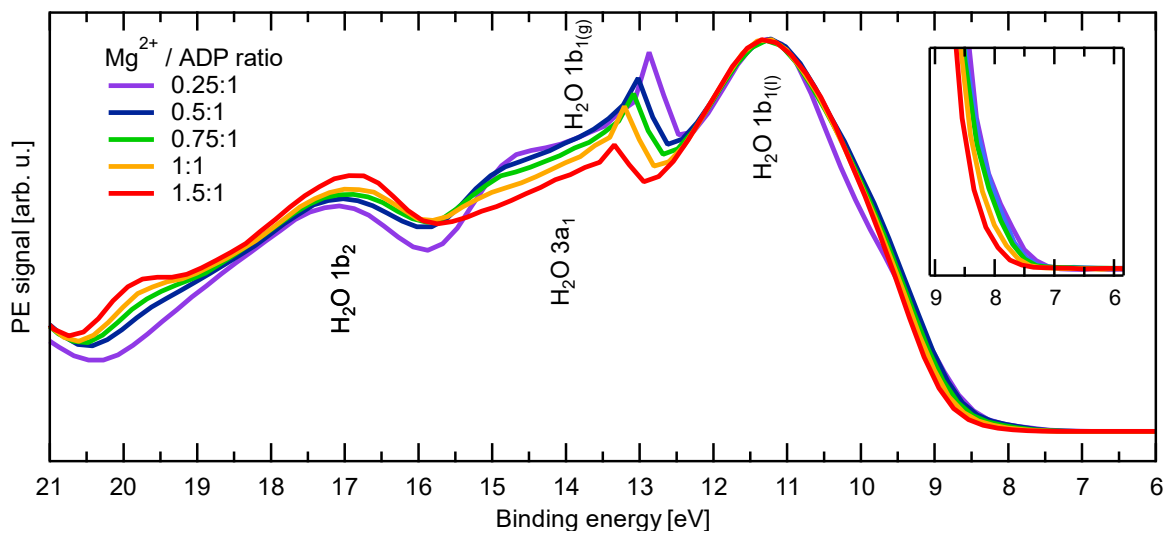

**Figure S4.** Valence and core-level (Mg 2p, Mg 2s, P 2p and P 2s) photoelectron spectra from 0.5 M ADP$_{(aq)}$ samples containing Mg$^{2+}$$_{(aq)}$ as a function of the Mg$^{2+}$/ADP concentration ratio recorded using a photon energy of 250 eV. The sample pH was 8.2, adjusted by addition of Tris. The BE scale was calibrated based on the liquid water solvent valence 1b$_1$ peak position, and spectral intensities are displayed to yield its same height. The figure inset highlights chemical shifts for the adenine unit, hinting to the presence of phosphate–Mg$^{2+}$–adenine interactions, as reported in Reference [9]; signal intensity is scaled analogous to the data shown in Figure 4(a) of the main manuscript.

**Linear baselines to the P 2s data presented in Figure 3e in the main text**

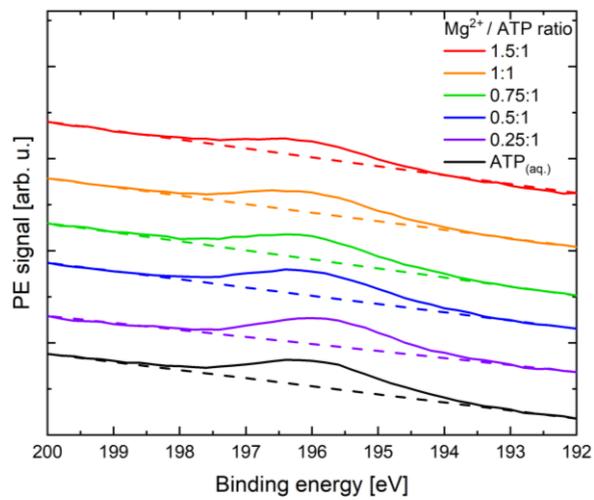

**Figure S5.** Linear baselines subtracted to the data presented in Figure 4e in the main text. The data are shown with a vertical offset for a better comparison.

**Data treatment to determine α-, β-, and γ-phosphate P 2s binding energies in ATP$_{(aq)}$**

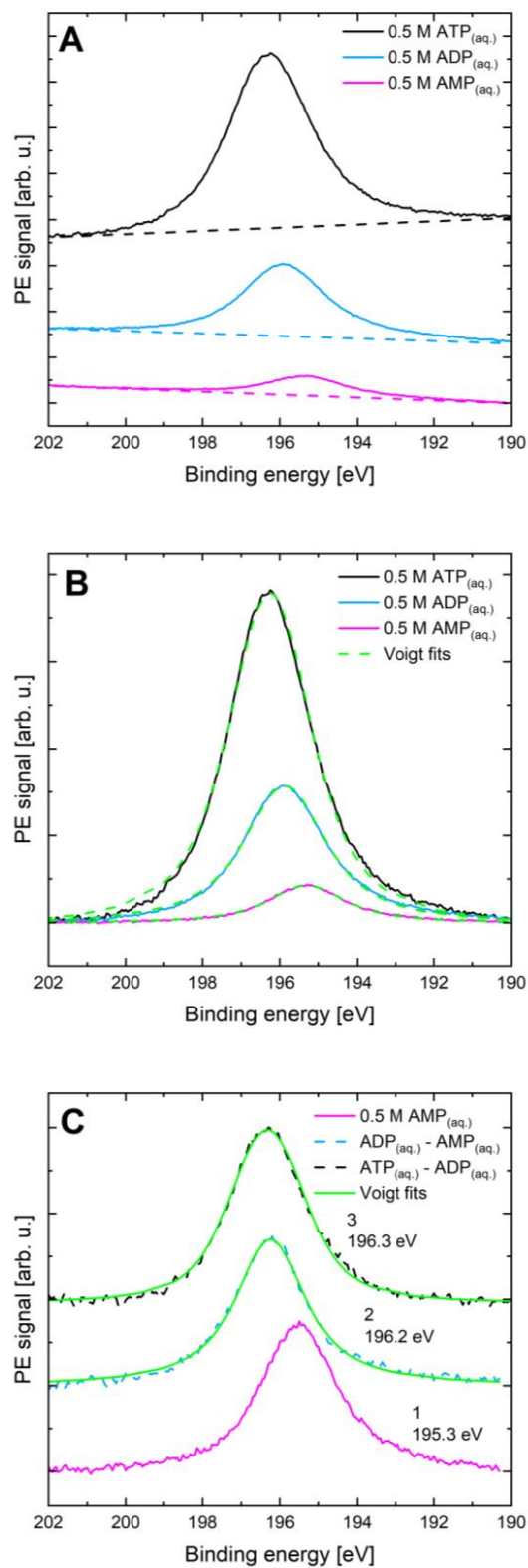

**Figure S6.** a) Linear baselines fit to and subtracted from the P 2s AMP$_{(aq)}$, ADP$_{(aq)}$ and ATP$_{(aq)}$ spectra. b) Voigt fits to the background-subtracted data performed to extract peak areas to re-scale each spectrum based on the number of phosphate units in each sample, as explained in the main text. c) Voigt fits to the AMP$_{(aq)}$ and the ADP$_{(aq)}$-AMP$_{(aq)}$ and ATP$_{(aq)}$-ADP$_{(aq)}$ difference spectra. The data are shown with a vertical offset for a better comparison. Peaks 1, 2 and 3 correspond to γ-, β-, and α-phosphate, as described in the main text.

**Intermolecular Coulombic decay (ICD) spectra of ATP$_{(aq)}$ solutions with dissolved Mg$^{2+}$**

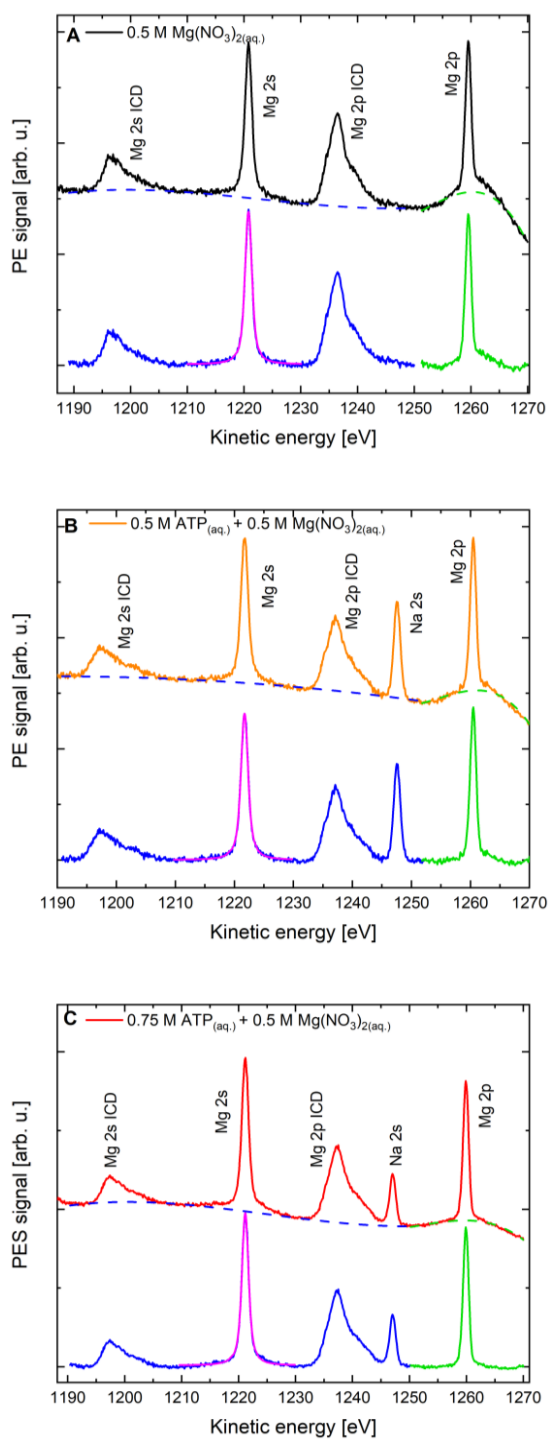

**Figure S7.** ICD spectra from a 0.5 M Mg(NO$_3$)$_{2(aq)}$ solution (panel a) and samples additionally containing ATP$_{(aq.)}$ at 1:1 and 1.5:1 Mg$^{2+}$/ATP concentration ratios (panels b and c, respectively). Cubic baselines were fit to and subtracted from different spectral regions – highlighted in green and blue dashed lines – to obtain the background-free data shown at the bottom of each plot and in Figure 8 in the main text (the use of a single baseline per spectrum did not allow the background to be removed completely, particularly in the vicinity of the Mg 2p peak). Voigt profile fits to the Mg 2s feature used to extract peak areas to normalize the signal intensity according to the Mg$^{2+}_{(aq.)}$ concentration, are shown in magenta.

**Sample input for calculations of P 2s binding energies using Q-Chem, version 6.0**

*$molecule*
*-2 1*
*C    7.224927   -0.822676   1.178578*
*C    6.400689    0.390773   1.648327*
*O    5.741657    0.889906   0.512408*
*C    6.357295    0.400165  -0.665258*
*C    6.764706   -1.015146  -0.271981*
*N    5.429865    0.126595   2.680502*
*C    4.426729   -0.802826   2.709145*
*C    3.760179   -0.587106   3.893105*
*N    4.331220    0.463841   4.589294*
*C    5.297970    0.850033   3.838275*
*N    4.132054   -1.736044   1.805532*
*C    3.098102   -2.463627   2.163904*
*N    2.364894   -2.366194   3.267622*
*C    2.673168   -1.429240   4.163609*
*N    1.957320   -1.353158   5.298273*
*C    5.383016    0.475700  -1.817277*
*O    5.034723    1.825785  -2.085047*
*P    5.533519    2.671333  -3.330859*
*O    7.086128    2.959546  -2.998166*
*P    7.796586    4.068300  -2.077932*
*O    9.114600    3.311294  -1.667383*
*P    10.678216   3.215316  -2.178902*
*O    10.922827   4.458632  -3.017769*
*O    7.777934   -1.506934  -1.105013*
*O    8.578676   -0.448645   1.231810*
*O    4.774182    3.950285  -3.325847*
*O    5.528548    1.824879  -4.556795*
*O    8.111996    5.247946  -2.943544*
*O    6.986746    4.312065  -0.857579*
*O    11.454269   3.203851  -0.883496*
*O    10.774878   1.951695  -2.989088*
*O    7.892952    3.910242   1.866903*
*O    8.886263    6.585860   1.873713*
*O    4.444713    3.630963   0.526843*
*O    6.631657    7.046811  -0.114316*
*O    2.995563    0.236236  -4.751444*
*O    7.395943   -1.171897  -7.894599*
*O    7.824774    0.477340  -5.580742*
*O    4.831772    2.698903  -7.194269*
*O    2.431757    3.795051  -1.615895*
*O    4.636346    6.771471  -2.335407*
*O    1.385435    1.259172  -2.503023*
*O    4.035588    5.163317  -5.855433*
*O    12.966150   4.476538  -5.001197*
*O    14.234390   4.236107  -1.054033*
*O    13.102625   0.908295  -0.257338*
*O    10.128784   2.142256   1.496512*
*O    11.223501   5.922243   0.264295*
*O    13.307154   0.620968  -3.152611*
*O    9.350816   -0.462957  -3.353000*
*O    10.034819   2.394841  -5.715182*
*O    15.023858   2.932689  -3.590595*
*O    10.321438   7.235626  -2.025491*
*O    6.733315    6.081292  -5.374094*
*O    9.472906    5.150173  -5.608173*
*Mg   9.851872    6.021003  -3.696028*

| | | | |
|---|---|---|---|
| H | 2.801346 | -3.237732 | 1.480314 |
| H | 2.035870 | -0.538121 | 5.866455 |
| H | 1.088072 | -1.840026 | 5.337770 |
| H | 5.967154 | 1.657172 | 4.057903 |
| H | 7.068832 | 1.137048 | 2.046875 |
| H | 7.057422 | -1.696537 | 1.791669 |
| H | 9.083442 | -1.045094 | 0.688867 |
| H | 5.903130 | -1.666458 | -0.293672 |
| H | 7.705345 | -2.449706 | -1.196279 |
| H | 7.241875 | 0.986102 | -0.886453 |
| H | 5.815846 | 0.010135 | -2.691753 |
| H | 4.465763 | -0.036247 | -1.564339 |
| H | 0.468714 | 1.380529 | -2.720858 |
| H | 3.103148 | 3.902189 | -2.288935 |
| H | 4.740231 | 5.856743 | -2.593883 |
| H | 4.110315 | 4.771779 | -4.985102 |
| H | 3.820253 | 0.716911 | -4.739126 |
| H | 5.122023 | 2.366727 | -6.342873 |
| H | 7.053876 | 0.926309 | -5.228815 |
| H | 6.666984 | -1.751152 | -7.707936 |
| H | 6.785323 | 6.124739 | -0.322195 |
| H | 5.173605 | 3.989064 | 0.023919 |
| H | 7.563655 | 3.969013 | 0.970381 |
| H | 8.535770 | 5.702781 | 2.004410 |
| H | 13.327405 | 3.942895 | -0.983119 |
| H | 11.243794 | 4.997076 | 0.018199 |
| H | 10.514336 | 2.459124 | 0.678530 |
| H | 12.526669 | 1.657321 | -0.424827 |
| H | 14.536007 | 2.115559 | -3.482834 |
| H | 12.451211 | 1.046945 | -3.232718 |
| H | 9.814979 | 0.346478 | -3.128514 |
| H | 10.344013 | 2.228645 | -4.821276 |
| H | 10.657318 | 6.801758 | -1.232143 |
| H | 12.299761 | 4.271847 | -4.340567 |
| H | 6.796653 | 5.660194 | -4.519793 |
| H | 8.570916 | 5.300176 | -5.884832 |
| H | 2.481603 | 0.582441 | -4.023542 |
| H | 7.518432 | -0.640673 | -7.106659 |
| H | 5.614821 | 2.787914 | -7.724618 |
| H | 8.259211 | 0.066565 | -4.830605 |
| H | 4.197890 | 4.439183 | -6.458247 |
| H | 4.757724 | 7.277308 | -3.130167 |
| H | 2.937496 | 3.704587 | -0.808888 |
| H | 1.692518 | 2.102843 | -2.164560 |
| H | 4.659113 | 2.704974 | 0.598189 |
| H | 8.656785 | 3.335200 | 1.816173 |
| H | 8.258972 | 7.017199 | 1.299901 |
| H | 5.957097 | 7.303932 | -0.737174 |
| H | 10.476687 | 6.050143 | 0.849382 |
| H | 9.690837 | 1.319313 | 1.296189 |
| H | 12.578179 | 0.273310 | 0.215853 |
| H | 14.571194 | 3.829778 | -1.850518 |
| H | 13.410962 | 0.503367 | -2.210164 |
| H | 14.479139 | 3.473358 | -4.159442 |
| H | 8.845222 | -0.714022 | -2.584581 |
| H | 9.265208 | 1.836569 | -5.818751 |
| H | 9.666441 | 7.863729 | -1.741839 |
| H | 12.703440 | 4.018541 | -5.791878 |
| H | 5.841635 | 5.917281 | -5.682034 |
| H | 9.630392 | 4.202328 | -5.678875 |

```
O    11.583857    6.941532   -4.587323
H    12.055758    7.491578   -3.970534
H    12.216189    6.299676   -4.913845
O     8.753769    7.733178   -4.382556
H     9.228149    8.347153   -4.933370
H     7.950080    7.504803   -4.851660
$end

$rem
METHOD HF
BASIS General
THRESH 12
MAX_SCF_CYCLES 129
solvent_method  pcm
PCM_PRINT      1
MEM_TOTAL  384000
MEM_STATIC 2000
$end

$basis
H 0
aug-cc-pVTZ
****
C 0
aug-cc-pVTZ
****
O 0
aug-cc-pVTZ
****
N 0
aug-cc-pVTZ
****
P 0
aug-cc-pCVTZ
****
Mg 0
aug-cc-pVTZ
****
$end

$pcm
THEORY         IEFPCM
RADII      Bondi
vdwScale       1.2
NonEquilibrium
$end

$solvent
Dielectric     78.39
OpticalDielectric 1.78
$end

@@@
$molecule
-1 2
READ
$end
$rem
METHOD HF
BASIS General
```

```
THRESH 12
MAX_SCF_CYCLES 129
unrestricted TRUE
mom_start   1
MOM_METHOD IMOM
scf_guess read
solvent_method  pcm
PCM_PRINT     1
MEM_TOTAL  384000
MEM_STATIC 2000
$end

$basis
H 0
aug-cc-pVTZ
****
C 0
aug-cc-pVTZ
****
O 0
aug-cc-pVTZ
****
N 0
aug-cc-pVTZ
****
P 0
aug-cc-pCVTZ
****
Mg 0
aug-cc-pVTZ
****
$end

$pcm
THEORY         IEFPCM
RADII     Bondi
vdwScale     1.2
StateSpecific     Marcus
$end

$solvent
Dielectric      78.39
OpticalDielectric 1.78
$end

$occupied
1:265
1:58 60:265
$end
```

**Cartesian coordinates of the structures used to calculate P 2s, Mg 2s, Mg 2p, and valence binding energies**

**ATP⁴⁻**
| | | | |
|---|---|---|---|
| O | 2.475207 | -3.395307 | 2.768565 |
| C | 3.116643 | -2.171512 | 3.011723 |
| C | 4.194717 | -2.008883 | 1.924691 |
| C | 4.177935 | -3.368016 | 1.207509 |
| C | 2.757811 | -3.863434 | 1.459706 |
| N | 3.592196 | -2.147772 | 4.372745 |
| C | 4.552282 | -2.919846 | 4.966112 |
| C | 4.579243 | -2.523986 | 6.283538 |
| N | 3.647530 | -1.526251 | 6.511357 |
| C | 3.094416 | -1.345473 | 5.366960 |
| N | 5.321334 | -3.857766 | 4.415860 |
| C | 6.144257 | -4.397138 | 5.286787 |
| N | 6.275861 | -4.116745 | 6.578892 |
| C | 5.501338 | -3.174173 | 7.114557 |
| N | 5.660240 | -2.861670 | 8.412372 |
| O | 3.762556 | -0.965352 | 1.096142 |
| O | 4.441559 | -3.287136 | -0.169428 |
| C | 2.609052 | -5.367908 | 1.432164 |
| O | 1.257592 | -5.732666 | 1.663907 |
| P | 0.321919 | -6.381603 | 0.554373 |
| O | -1.008187 | -6.594551 | 1.188953 |
| O | 0.993617 | -7.558026 | -0.068742 |
| O | 0.260261 | -5.166480 | -0.489728 |
| P | -0.375405 | -4.956620 | -1.955765 |
| O | -1.253088 | -6.112303 | -2.277146 |
| O | -0.980812 | -3.593681 | -1.953029 |
| O | 0.956906 | -4.993922 | -2.814477 |
| P | 1.431486 | -4.531127 | -4.329516 |
| O | 1.507941 | -3.022919 | -4.302705 |
| O | 2.798210 | -5.183288 | -4.453890 |
| O | 0.402891 | -5.085745 | -5.285297 |
| O | -1.950813 | -3.896050 | -6.208541 |
| O | 3.369355 | -1.764794 | -2.602713 |
| O | -4.815969 | -7.715183 | -0.363696 |
| O | 1.171972 | -6.227337 | -7.798200 |
| O | -0.066080 | -0.677557 | -4.447338 |
| O | -0.329941 | -9.806943 | -1.223062 |
| O | -2.513122 | -2.186627 | -3.927615 |
| O | 3.926770 | -5.658181 | -7.051304 |
| O | 3.362062 | -8.234585 | -1.563689 |
| O | 4.442487 | -5.642485 | -2.161818 |
| O | -1.076998 | -6.252734 | 4.049765 |
| O | 1.166574 | -7.970530 | 4.602655 |
| O | 3.029755 | -1.813375 | -6.445368 |
| O | -3.698626 | -5.550410 | 1.249475 |
| O | -3.269544 | -3.086843 | -0.123536 |
| O | -1.664018 | -1.818526 | 1.950515 |
| O | 0.009658 | -1.377793 | -0.344203 |
| O | -0.290018 | -3.447608 | 3.925356 |
| O | 1.419191 | 0.382854 | -2.192692 |
| O | 2.864183 | -8.078403 | -4.448286 |
| O | 1.626454 | -9.012748 | -6.928543 |
| O | -2.451152 | -9.059588 | 0.636424 |
| O | 1.975078 | -9.354465 | 2.127703 |
| O | 5.013537 | -3.153315 | -4.641851 |
| H | 6.796334 | -5.165080 | 4.913079 |
| H | 4.953714 | -2.328459 | 8.870541 |
| H | 6.179552 | -3.492127 | 8.984392 |
| H | 2.313712 | -0.641374 | 5.161558 |
| H | 2.412108 | -1.358978 | 2.924072 |
| H | 5.164218 | -1.778983 | 2.344231 |
| H | 4.335341 | -0.895551 | 0.341894 |
| H | 4.888333 | -4.029205 | 1.684220 |
| H | 5.379139 | -3.215837 | -0.314038 |
| H | 2.076008 | -3.407579 | 0.752399 |
| H | 2.952773 | -5.750967 | 0.482470 |
| H | 3.198696 | -5.813628 | 2.220668 |
| H | -1.118061 | -6.430238 | 3.109629 |
| H | 0.592011 | -3.503467 | 3.566691 |
| H | -2.789173 | -5.849289 | 1.250759 |
| H | -1.983174 | -8.265615 | 0.896974 |
| H | 0.912742 | -8.581517 | 5.284363 |
| H | 0.156337 | -9.044528 | -0.902593 |
| H | 2.563180 | -8.034850 | -1.076572 |
| H | 1.681047 | -8.824445 | 1.390148 |
| H | -2.558492 | -3.269788 | -0.736625 |
| H | -2.306234 | -2.255737 | 1.391350 |
| H | -0.250710 | -2.101641 | -0.914611 |
| H | -2.058230 | -2.681111 | -3.244459 |
| H | 2.427856 | -2.180773 | -5.796466 |
| H | 0.443952 | -1.491402 | -4.447315 |
| H | 0.872055 | 0.172392 | -2.950488 |
| H | 2.701292 | -2.239892 | -3.109062 |
| H | 3.582167 | -5.471233 | -6.174368 |
| H | 4.356278 | -3.847058 | -4.599979 |
| H | 3.853624 | -5.447558 | -2.894940 |
| H | 2.809894 | -7.121409 | -4.477703 |
| H | -1.130256 | -4.294734 | -5.897970 |
| H | 0.850963 | -5.829439 | -6.985799 |
| H | -4.600024 | -7.330510 | -1.213263 |
| H | 2.024280 | -8.829576 | -6.079103 |
| H | 0.418860 | -7.381479 | 4.481141 |
| H | 1.696649 | -8.887439 | 2.913933 |
| H | -0.512242 | -9.635718 | -2.139404 |
| H | 3.813274 | -7.394948 | -1.651559 |
| H | -1.890087 | -9.460721 | -0.025083 |
| H | -4.181022 | -6.211003 | 0.753530 |
| H | -0.795029 | -2.946504 | 3.283477 |
| H | -0.841854 | -5.326339 | 4.107245 |
| H | -0.995615 | -1.524300 | 1.333096 |
| H | 0.467357 | -0.747213 | -0.897957 |
| H | -2.431064 | -2.722281 | -4.716296 |
| H | -3.493661 | -3.928017 | 0.276237 |
| H | -0.960771 | -0.960535 | -4.265626 |
| H | 2.180614 | -0.189848 | -2.274852 |
| H | 3.508452 | -2.259853 | -1.800260 |
| H | 3.885953 | -2.095804 | -6.137286 |
| H | 4.766434 | -2.560258 | -3.936695 |
| H | 4.327454 | -4.852197 | -7.354662 |
| H | 4.331957 | -4.934790 | -1.534834 |
| H | 2.952167 | -8.297053 | -3.521721 |
| H | -1.722774 | -3.391847 | -6.980429 |
| H | 2.106160 | -6.027763 | -7.802235 |
| H | -4.075683 | -8.275210 | -0.128717 |
| H | 1.397894 | -8.155122 | -7.283145 |
| H | -3.085445 | -6.169534 | -2.604806 |
| H | -0.462198 | -6.833069 | -5.070542 |
| O | -0.959272 | -7.570695 | -4.717995 |
| H | -1.139700 | -7.282010 | -3.825103 |
| O | -4.028635 | -6.285738 | -2.748432 |
| H | -4.116345 | -6.689773 | -3.603566 |

**[Mg(ATP)$_2$]$^{6-}$**

| | | | |
|---|---|---|---|
| C | 15.122721 | 5.138014 | -3.504118 |
| N | 13.906459 | 4.566719 | -3.274397 |
| C | 12.978463 | 5.501577 | -3.646856 |
| N | 13.492853 | 6.597815 | -4.084327 |
| C | 14.857403 | 6.392645 | -4.002849 |
| C | 13.695646 | 3.225794 | -2.745279 |
| C | 13.655959 | 3.154269 | -1.208974 |
| C | 12.152796 | 3.108730 | -0.923158 |
| C | 11.660327 | 2.306619 | -2.111870 |
| O | 12.463419 | 2.756673 | -3.192572 |
| C | 10.236685 | 2.530503 | -2.511529 |
| O | 9.447062 | 1.864144 | -1.564477 |
| P | 8.058127 | 2.398362 | -1.017189 |
| O | 8.143991 | 3.826765 | -0.643504 |
| O | 11.889457 | 2.575883 | 0.357547 |
| O | 14.223319 | 1.953109 | -0.763739 |
| O | 7.629475 | 1.416794 | 0.008996 |
| Mg | 7.277821 | -0.619513 | 0.235947 |
| O | 6.542116 | -0.761625 | -1.652732 |
| P | 5.526419 | -0.766091 | -2.760918 |
| O | 6.008052 | -1.474981 | -4.011919 |
| O | 7.011058 | -2.731423 | 0.507198 |
| P | 6.142188 | -3.593151 | 1.357725 |
| O | 6.182023 | -5.060980 | 1.166976 |
| O | 4.650139 | -3.038669 | 1.205605 |
| C | 3.473581 | -3.815468 | 1.414581 |
| C | 3.243752 | -4.160418 | 2.868156 |
| C | 2.066049 | -5.097165 | 3.119522 |
| C | 1.738992 | -4.831982 | 4.604703 |
| C | 2.335936 | -3.422040 | 4.835981 |
| O | 2.891889 | -3.009358 | 3.612834 |
| N | 1.414841 | -2.415669 | 5.295979 |
| C | 0.254853 | -1.980160 | 4.713765 |
| C | -0.177077 | -0.933963 | 5.495298 |
| N | 0.699507 | -0.717399 | 6.545268 |
| C | 1.611613 | -1.606421 | 6.381914 |
| N | -0.354311 | -2.447602 | 3.625852 |
| C | -1.456730 | -1.785844 | 3.353718 |
| N | -1.987882 | -0.763065 | 4.014993 |
| C | -1.367394 | -0.307689 | 5.102385 |

```
N   -1.923525    0.701920    5.794503
O    2.280263   -5.763823    5.489029
O    2.423516   -6.413040    2.818043
O    6.383011   -3.241808    2.909289
P    7.634561   -3.192847    3.911696
O    7.272420   -3.979698    5.131771
O    8.071605   -0.535251    2.135405
P    7.693606   -0.227748    3.556581
O    6.277872    0.307161    3.696041
O    8.718909    0.561546    4.332315
O    7.667652   -1.678156    4.351118
O    8.880105   -3.591552    3.196954
O    7.163961    2.277224   -2.346436
P    5.576165    2.216176   -2.621152
O    4.823489    2.344435   -1.354096
O    5.407556    0.798225   -3.287956
O    5.288968    3.209570   -3.708924
O    4.129284   -1.143834   -2.343945
O   13.453017    2.315472    2.808501
O    9.425963   -0.296207    6.922565
O    3.337847   -0.060487    3.553376
O   10.796150   -2.290349    1.542425
O   11.269579   -4.628879    0.020866
O   11.602513    0.765424    4.373071
O    5.873058    3.150462    3.397641
O    7.671667   -5.599681   -3.347842
O    5.195813    0.461980    6.374736
O    2.695263    1.811472    7.068843
O   12.700419   -1.806246    3.665144
O    7.032244   -0.355753    8.525155
O    8.904493   -3.686264   -1.477933
O   11.222240   -3.885209    5.018588
O    9.380493   -3.226530    7.173552
O    9.911080   -6.053320    2.210080
O    5.585024   -2.522216    7.076878
O    8.708855    3.396166    3.886662
O    7.958028   -7.184363    0.333688
O    5.233082   -7.296355    2.738408
O    5.171714   -6.058666    5.365088
O    8.777708   -6.637859    5.073210
H   -2.002917   -2.107345    2.486061
H   -1.368880    1.190363    6.463363
H   -2.644196    1.229116    5.350426
H    2.466447   -1.735190    7.013407
H    3.109003   -3.488982    5.586823
H    0.675322   -4.838949    4.778696
H    3.231766   -5.832860    5.401951
H    1.222619   -4.786651    2.514473
H    1.671923   -6.988752    2.896904
H    4.140100   -4.595735    3.289555
H    3.543228   -4.726858    0.839419
H    2.658389   -3.215312    1.039463
H    6.894418   -5.151112   -3.680245
H    8.239205   -3.558572   -0.800139
H    7.309273   -6.492084    0.443350
H    5.599118   -6.589954    2.205269
```

| | | | |
|---|---|---|---|
| H | 10.521191 | -3.883869 | 4.368628 |
| H | 9.489560 | -5.233948 | 2.474692 |
| H | 11.291042 | -3.802824 | 0.510617 |
| H | 10.088979 | -2.629774 | 2.096079 |
| H | 8.640149 | -3.508179 | 6.639479 |
| H | 6.118903 | -2.981935 | 6.429409 |
| H | 5.806243 | -5.343355 | 5.334948 |
| H | 8.323192 | -5.798207 | 5.096767 |
| H | 10.652971 | 0.678913 | 4.267172 |
| H | 8.808623 | 2.453470 | 4.046224 |
| H | 7.880934 | -0.286171 | 8.089511 |
| H | 9.180961 | -0.005652 | 6.036329 |
| H | 4.267426 | 0.040041 | 3.744427 |
| H | 2.007647 | 1.166513 | 6.933982 |
| H | 5.655510 | 0.485307 | 5.533309 |
| H | 5.941906 | 2.195097 | 3.439275 |
| H | 13.096563 | 2.389663 | 1.924657 |
| H | 12.207221 | -1.902899 | 2.851275 |
| H | 8.079500 | -4.993869 | -2.731702 |
| H | 4.287771 | -7.190708 | 2.668681 |
| H | 9.650418 | -4.098528 | -1.044069 |
| H | 8.643849 | -6.953484 | 0.956939 |
| H | 10.378742 | -1.658413 | 0.961722 |
| H | 12.108829 | -4.710725 | -0.417305 |
| H | 10.537239 | -5.797659 | 1.535887 |
| H | 10.777370 | -3.679279 | 5.839241 |
| H | 9.390331 | -2.271661 | 7.118297 |
| H | 5.250605 | -1.746240 | 6.636125 |
| H | 5.313004 | -6.554580 | 4.559313 |
| H | 9.249930 | -6.650342 | 4.247052 |
| H | 7.778891 | 3.499760 | 3.689979 |
| H | 6.694974 | -1.216141 | 8.284260 |
| H | 10.236271 | 0.151458 | 7.136227 |
| H | 11.970934 | -0.102727 | 4.202663 |
| H | 5.862767 | 0.455737 | 7.058054 |
| H | 3.509382 | 1.383351 | 6.806049 |
| H | 3.194736 | -0.999391 | 3.474705 |
| H | 5.492138 | 3.362202 | 2.547132 |
| H | 12.783967 | 1.859401 | 3.320709 |
| H | 12.317237 | -2.456972 | 4.252614 |
| H | 10.059966 | 2.101246 | -3.487768 |
| H | 9.993399 | 3.583286 | -2.540051 |
| H | 11.759206 | 4.117314 | -0.998480 |
| H | 14.145408 | 4.004835 | -0.754462 |
| H | 14.495834 | 2.614531 | -3.132882 |
| H | 11.838995 | 1.250632 | -1.945347 |
| H | 11.168941 | 3.089791 | 0.785453 |
| H | 15.170804 | 2.009521 | -0.795295 |
| H | 11.931646 | 5.306381 | -3.580631 |
| N | 16.331728 | 4.614211 | -3.298705 |
| C | 17.292568 | 5.443459 | -3.639176 |
| N | 17.183460 | 6.672802 | -4.131665 |
| C | 15.967360 | 7.183678 | -4.326565 |
| H | 18.297853 | 5.088404 | -3.505079 |
| N | 15.858576 | 8.441332 | -4.788328 |
| H | 16.669169 | 8.867457 | -5.182710 |
| H | 14.978680 | 8.755342 | -5.135152 |
| O | 9.279387 | -0.967748 | -0.578293 |
| H | 9.296387 | -1.710180 | -1.181603 |
| H | 9.541796 | -0.192303 | -1.069819 |
| H | 2.325789 | 3.294184 | -6.149083 |
| H | 0.885710 | 0.625091 | -1.431151 |
| H | 3.268509 | -2.648227 | -2.847018 |
| H | 5.651851 | -3.256993 | -4.332366 |
| H | 5.130943 | -1.069390 | -5.575432 |
| H | 7.629637 | -0.934966 | -4.711083 |
| H | 1.772295 | -4.619872 | -2.016357 |
| H | 5.078375 | 5.113043 | -3.161387 |
| H | 6.873340 | 4.031126 | -4.569015 |
| H | 3.359321 | 3.259062 | -3.968764 |
| H | 3.473369 | 1.179684 | 0.593260 |
| H | 4.657122 | 3.392850 | 0.148916 |
| H | 9.244449 | 4.239346 | 0.805459 |
| H | 6.921926 | 5.244031 | -0.359312 |
| O | 5.359108 | -0.231104 | 1.070584 |
| H | 4.817547 | -1.016001 | 1.101030 |
| H | 5.542164 | 0.006090 | 1.984167 |
| H | 9.287225 | 6.308592 | -3.018087 |
| H | 10.155320 | 3.242373 | -8.631937 |
| H | 5.198593 | 2.339493 | -5.534371 |
| H | 7.126298 | 1.847552 | -6.663620 |
| H | 13.890636 | 4.036335 | 3.510037 |
| H | 2.660434 | -0.075290 | -2.883271 |
| O | 14.157138 | 4.892007 | 3.850850 |
| H | 13.356159 | 5.390936 | 3.956311 |
| O | 9.943975 | 4.296103 | 1.444816 |
| H | 9.583821 | 3.970587 | 2.271228 |
| O | 6.257400 | 5.921835 | -0.219878 |
| H | 5.577839 | 5.496061 | 0.297064 |
| O | 4.436953 | 3.901070 | 0.933899 |
| H | 3.618606 | 3.517986 | 1.236609 |
| O | 1.238452 | -5.260748 | -1.547845 |
| H | 1.012676 | -4.855234 | -0.719768 |
| O | 9.416528 | 6.289534 | -3.958971 |
| H | 8.831081 | 5.611751 | -4.292506 |
| O | 9.639252 | 2.486624 | -8.885695 |
| H | 9.119353 | 2.253960 | -8.115737 |
| O | 8.077183 | 1.832393 | -6.550585 |
| H | 8.264880 | 1.031211 | -6.057745 |
| O | 8.464824 | -0.643217 | -5.094734 |
| H | 9.071753 | -0.565787 | -4.368010 |
| O | 5.010833 | 6.023032 | -2.877133 |
| H | 5.415714 | 6.051349 | -2.011373 |
| O | 7.626788 | 4.338730 | -5.067684 |
| H | 7.966797 | 3.560476 | -5.504564 |
| O | 2.421581 | 3.130841 | -4.123140 |
| H | 1.972521 | 3.762477 | -3.572943 |
| O | 2.719646 | 1.401988 | 1.131653 |
| H | 2.858149 | 0.920103 | 1.948224 |
| O | 0.448099 | 0.757347 | -0.589911 |
| H | 1.146564 | 0.954536 | 0.032205 |
| O | 5.400157 | -4.179386 | -4.439086 |
| H | 4.539110 | -4.231532 | -4.031733 |
| O | 4.714992 | -0.801526 | -6.402789 |
| H | 3.802607 | -1.057091 | -6.335677 |
| O | 2.827070 | -3.451517 | -3.154714 |
| H | 2.291041 | -3.185862 | -3.893235 |
| O | 5.191203 | 2.025902 | -6.438664 |
| H | 4.966458 | 1.094410 | -6.396698 |
| O | 1.846360 | 0.381491 | -3.112698 |
| H | 2.093385 | 1.248825 | -3.426677 |
| O | 2.581254 | 3.242391 | -7.067314 |
| H | 3.453200 | 2.852624 | -7.042568 |

**[Mg-ATP]$^{2-}$ (Mg$^{2+}$ bound to α-, β-, and γ-phosphate)**

| | | | |
|---|---|---|---|
| C | 0.954635 | -3.123957 | 3.443092 |
| C | 1.610854 | -1.889433 | 3.533673 |
| C | 2.789449 | -1.766177 | 2.834651 |
| N | 3.356160 | -2.711252 | 2.086101 |
| C | 2.654092 | -3.822032 | 2.074538 |
| N | 1.508199 | -4.078531 | 2.695855 |
| N | 1.315758 | -0.714236 | 4.201936 |
| C | 2.287604 | 0.072025 | 3.904475 |
| N | 3.229610 | -0.495530 | 3.083724 |
| C | 4.391895 | 0.185018 | 2.568644 |
| O | 4.242835 | 0.366270 | 1.182247 |
| C | 5.473575 | 0.158313 | 0.513960 |
| C | 6.155354 | -0.917085 | 1.351917 |
| C | 5.737815 | -0.535146 | 2.777581 |
| O | 6.619309 | 0.403858 | 3.330351 |
| O | 7.543763 | -0.887231 | 1.156396 |
| C | 5.220621 | -0.256326 | -0.917086 |
| O | 4.847572 | 0.858966 | -1.721345 |
| P | 5.789484 | 1.577778 | -2.784268 |
| O | 6.984715 | 0.754435 | -3.087353 |
| N | -0.184353 | -3.394308 | 4.104090 |
| O | 4.870864 | 1.946999 | -3.908973 |
| O | 6.278113 | 2.897527 | -2.038129 |
| P | 5.823876 | 4.448014 | -1.974217 |
| O | 6.846183 | 5.111237 | -1.112622 |
| O | 5.644011 | 4.967985 | -3.356215 |
| O | 4.471795 | 4.388940 | -1.167305 |
| P | 2.848690 | 4.363266 | -1.480705 |
| O | 2.273622 | 3.883838 | -0.170176 |
| O | 2.473143 | 5.777729 | -1.840163 |
| O | 2.660675 | 3.386401 | -2.615988 |
| O | 4.226696 | 4.755847 | -5.816853 |
| O | 6.428076 | 4.179170 | -7.531633 |
| O | 5.797239 | 7.788283 | -3.992978 |
| O | 9.037713 | 0.648255 | -0.981255 |
| O | 5.813876 | -1.040892 | -5.171126 |
| O | 5.609770 | 1.480859 | -6.719840 |
| O | 9.141228 | 2.015070 | -4.648081 |
| O | 9.046779 | 3.412046 | 0.024391 |
| O | 9.244509 | 6.024718 | -2.708953 |
| O | 6.329650 | 8.091998 | -1.109363 |
| O | 6.278299 | 5.336834 | 1.747569 |

```
O     2.947777    1.272365   -7.804559
O     3.140877   -0.194173   -4.501277
O     7.966584    4.676847   -5.088271
Mg    2.867314    1.752391   -3.688918
H     3.051403   -4.630130    1.488144
H    -0.714854   -2.639517    4.481100
H    -0.697209   -4.207287    3.839005
H     2.393032    1.084118    4.242168
H     4.427132    1.141974    3.064286
H     5.644278   -1.392764    3.429316
H     7.512979    0.142518    3.136512
H     5.755873   -1.890504    1.106083
H     7.915831   -1.752578    1.281506
H     6.063134    1.067817    0.540104
H     6.104656   -0.719004   -1.327820
H     4.401396   -0.958707   -0.960809
H     6.342901   -0.625570   -4.491640
H     5.404007    1.741385   -5.822270
H     8.465148    1.553130   -4.156892
H     8.420425    0.650216   -1.711753
H     5.773822    6.843921   -3.838579
H     7.174615    4.740860   -4.553539
H     5.625908    4.489096   -7.106086
H     4.628638    4.723901   -4.946163
H     6.574670    7.170756   -1.046940
H     6.465825    5.258845    0.812505
H     8.320828    3.889258   -0.370453
H     8.508768    5.763186   -2.159212
H     1.707619    6.362084   -3.443250
H     0.730087    6.483577   -1.214766
H     3.865125    7.650362    1.361416
H     3.124921    7.258201   -0.976322
H     3.871636    1.252867   -7.546872
H     4.009729   -0.547629   -4.711939
H     1.825137    2.132131    0.000714
H     3.342771    3.390641    5.372239
H     3.065684    4.255242    1.474931
H     0.630110    4.277967    0.628412
H    -0.686783    4.208028   -5.020045
H     3.757463    6.661855   -6.012344
H     5.931473   -0.487630   -5.938829
H     8.613946    0.147937   -0.288697
H     6.010330    2.247528   -7.125850
H     8.803038    2.900373   -4.771067
H     2.622088   -0.902624   -4.136851
H     2.692880    0.368674   -7.950677
H     3.547236    4.088126   -5.810015
H     6.464123    4.590023   -8.387786
H     7.656990    4.587318   -5.987826
H     5.989805    8.155332   -3.132621
H     5.407970    8.126914   -0.856794
H     5.511906    4.790034    1.902835
H     9.016749    2.527743   -0.336788
H     9.045734    5.676645   -3.573110
O    -0.110281    6.781431   -0.867759
H    -0.332542    6.139911   -0.196873
```

| | | | |
|---|---|---|---|
| O | 4.153383 | 7.361749 | 2.226547 |
| H | 5.013855 | 6.971798 | 2.087350 |
| O | 3.482357 | 8.026677 | -0.515964 |
| H | 2.942277 | 8.768045 | -0.763970 |
| O | 1.272554 | 6.691656 | -4.233959 |
| H | 1.965647 | 7.068193 | -4.774832 |
| O | 3.463097 | 4.497936 | 2.316446 |
| H | 3.425215 | 5.451950 | 2.355871 |
| O | 2.579862 | 3.373276 | 4.807144 |
| H | 2.868759 | 3.734246 | 3.967829 |
| O | 1.598700 | 1.196017 | 0.029647 |
| H | 2.375341 | 0.766625 | 0.377667 |
| O | -0.205282 | 4.547645 | 1.019815 |
| H | -0.825788 | 3.856791 | 0.819887 |
| O | 0.106611 | 4.426524 | -5.495140 |
| H | 0.468923 | 5.208172 | -5.067267 |
| O | 3.542788 | 7.578891 | -5.845920 |
| H | 4.210357 | 7.871481 | -5.227066 |
| O | 2.129293 | 2.516227 | -5.431968 |
| H | 2.226428 | 2.064690 | -6.276298 |
| H | 1.378198 | 3.119680 | -5.480290 |
| O | 1.302677 | 0.762907 | -2.684577 |
| H | 0.413279 | 0.975813 | -2.949080 |
| H | 1.338145 | 0.870184 | -1.723505 |

**[Mg-ATP]$^{2-}$ (Mg$^{2+}$ bound to β- and γ-phosphate)**

| | | | |
|---|---|---|---|
| C | 7.224927 | -0.822676 | 1.178578 |
| C | 6.400689 | 0.390773 | 1.648327 |
| O | 5.741657 | 0.889906 | 0.512408 |
| C | 6.357295 | 0.400165 | -0.665258 |
| C | 6.764706 | -1.015146 | -0.271981 |
| N | 5.429865 | 0.126595 | 2.680502 |
| C | 4.426729 | -0.802826 | 2.709145 |
| C | 3.760179 | -0.587106 | 3.893105 |
| N | 4.331220 | 0.463841 | 4.589294 |
| C | 5.297970 | 0.850033 | 3.838275 |
| N | 4.132054 | -1.736044 | 1.805532 |
| C | 3.098102 | -2.463627 | 2.163904 |
| N | 2.364894 | -2.366194 | 3.267622 |
| C | 2.673168 | -1.429240 | 4.163609 |
| N | 1.957320 | -1.353158 | 5.298273 |
| C | 5.383016 | 0.475700 | -1.817277 |
| O | 5.034723 | 1.825785 | -2.085047 |
| P | 5.533519 | 2.671333 | -3.330859 |
| O | 7.086128 | 2.959546 | -2.998166 |
| P | 7.796586 | 4.068300 | -2.077932 |
| O | 9.114600 | 3.311294 | -1.667383 |
| P | 10.678216 | 3.215316 | -2.178902 |
| O | 10.922827 | 4.458632 | -3.017769 |
| O | 7.777934 | -1.506934 | -1.105013 |
| O | 8.578676 | -0.448645 | 1.231810 |
| O | 4.774182 | 3.950285 | -3.325847 |
| O | 5.528548 | 1.824879 | -4.556795 |
| O | 8.111996 | 5.247946 | -2.943544 |
| O | 6.986746 | 4.312065 | -0.857579 |
| O | 11.454269 | 3.203851 | -0.883496 |
| O | 10.774878 | 1.951695 | -2.989088 |
| O | 7.892952 | 3.910242 | 1.866903 |
| O | 8.886263 | 6.585860 | 1.873713 |
| O | 4.444713 | 3.630963 | 0.526843 |
| O | 6.631657 | 7.046811 | -0.114316 |
| O | 2.995563 | 0.236236 | -4.751444 |
| O | 7.395943 | -1.171897 | -7.894599 |
| O | 7.824774 | 0.477340 | -5.580742 |
| O | 4.831772 | 2.698903 | -7.194269 |
| O | 2.431757 | 3.795051 | -1.615895 |
| O | 4.636346 | 6.771471 | -2.335407 |
| O | 1.385435 | 1.259172 | -2.503023 |
| O | 4.035588 | 5.163317 | -5.855433 |
| O | 12.966150 | 4.476538 | -5.001197 |
| O | 14.234390 | 4.236107 | -1.054033 |
| O | 13.102625 | 0.908295 | -0.257338 |
| O | 10.128784 | 2.142256 | 1.496512 |
| O | 11.223501 | 5.922243 | 0.264295 |
| O | 13.307154 | 0.620968 | -3.152611 |
| O | 9.350816 | -0.462957 | -3.353000 |
| O | 10.034819 | 2.394841 | -5.715182 |
| O | 15.023858 | 2.932689 | -3.590595 |
| O | 10.321438 | 7.235626 | -2.025491 |
| O | 6.733315 | 6.081292 | -5.374094 |
| O | 9.472906 | 5.150173 | -5.608173 |
| Mg | 9.851872 | 6.021003 | -3.696028 |
| H | 2.801346 | -3.237732 | 1.480314 |
| H | 2.035870 | -0.538121 | 5.866455 |
| H | 1.088072 | -1.840026 | 5.337770 |
| H | 5.967154 | 1.657172 | 4.057903 |
| H | 7.068832 | 1.137048 | 2.046875 |
| H | 7.057422 | -1.696537 | 1.791669 |
| H | 9.083442 | -1.045094 | 0.688867 |
| H | 5.903130 | -1.666458 | -0.293672 |
| H | 7.705345 | -2.449706 | -1.196279 |
| H | 7.241875 | 0.986102 | -0.886453 |
| H | 5.815846 | 0.010135 | -2.691753 |
| H | 4.465763 | -0.036247 | -1.564339 |
| H | 0.468714 | 1.380529 | -2.720858 |
| H | 3.103148 | 3.902189 | -2.288935 |
| H | 4.740231 | 5.856743 | -2.593883 |
| H | 4.110315 | 4.771779 | -4.985102 |
| H | 3.820253 | 0.716911 | -4.739126 |
| H | 5.122023 | 2.366727 | -6.342873 |
| H | 7.053876 | 0.926309 | -5.228815 |
| H | 6.666984 | -1.751152 | -7.707936 |
| H | 6.785323 | 6.124739 | -0.322195 |
| H | 5.173605 | 3.989064 | 0.023919 |
| H | 7.563655 | 3.969013 | 0.970381 |
| H | 8.535770 | 5.702781 | 2.004410 |
| H | 13.327405 | 3.942895 | -0.983119 |
| H | 11.243794 | 4.997076 | 0.018199 |
| H | 10.514336 | 2.459124 | 0.678530 |
| H | 12.526669 | 1.657321 | -0.424827 |
| H | 14.536007 | 2.115559 | -3.482834 |
| H | 12.451211 | 1.046945 | -3.232718 |
| H | 9.814979 | 0.346478 | -3.128514 |
| H | 10.344013 | 2.228645 | -4.821276 |
| H | 10.657318 | 6.801758 | -1.232143 |
| H | 12.299761 | 4.271847 | -4.340567 |
| H | 6.796653 | 5.660194 | -4.519793 |
| H | 8.570916 | 5.300176 | -5.884832 |
| H | 2.481603 | 0.582441 | -4.023542 |
| H | 7.518432 | -0.640673 | -7.106659 |
| H | 5.614821 | 2.787914 | -7.724618 |
| H | 8.259211 | 0.066565 | -4.830605 |
| H | 4.197890 | 4.439183 | -6.458247 |
| H | 4.757724 | 7.277308 | -3.130167 |
| H | 2.937496 | 3.704587 | -0.808888 |
| H | 1.692518 | 2.102843 | -2.164560 |
| H | 4.659113 | 2.704974 | 0.598189 |
| H | 8.656785 | 3.335200 | 1.816173 |
| H | 8.258972 | 7.017199 | 1.299901 |
| H | 5.957097 | 7.303932 | -0.737174 |
| H | 10.476687 | 6.050143 | 0.849382 |
| H | 9.690837 | 1.319313 | 1.296189 |
| H | 12.578179 | 0.273310 | 0.215853 |
| H | 14.571194 | 3.829778 | -1.850518 |
| H | 13.410962 | 0.503367 | -2.210164 |
| H | 14.479139 | 3.473358 | -4.159442 |
| H | 8.845222 | -0.714022 | -2.584581 |
| H | 9.265208 | 1.836569 | -5.818751 |
| H | 9.666441 | 7.863729 | -1.741839 |
| H | 12.703440 | 4.018541 | -5.791878 |
| H | 5.841635 | 5.917281 | -5.682034 |
| H | 9.630392 | 4.202328 | -5.678875 |
| O | 11.583857 | 6.941532 | -4.587323 |
| H | 12.055758 | 7.491578 | -3.970534 |
| H | 12.216189 | 6.299676 | -4.913845 |
| O | 8.753769 | 7.733178 | -4.382556 |
| H | 9.228149 | 8.347153 | -4.933370 |
| H | 7.950080 | 7.504803 | -4.851660 |

**Mg$_2$ATP**

| | | | |
|---|---:|---:|---:|
| C | 4.804353 | -2.044697 | 2.701936 |
| C | 3.876445 | -0.891776 | 2.280688 |
| O | 4.622018 | -0.061753 | 1.423674 |
| C | 6.001809 | -0.379617 | 1.484242 |
| C | 6.021542 | -1.875544 | 1.783454 |
| N | 2.652832 | -1.277726 | 1.621047 |
| C | 2.474243 | -2.055614 | 0.510115 |
| C | 1.117379 | -2.065474 | 0.282912 |
| N | 0.458978 | -1.302476 | 1.231278 |
| C | 1.398245 | -0.861098 | 1.986650 |
| N | 3.392489 | -2.690404 | -0.216019 |
| C | 2.860173 | -3.355109 | -1.216816 |
| N | 1.578346 | -3.451237 | -1.552825 |
| C | 0.670651 | -2.813736 | -0.814591 |
| N | -0.629742 | -2.942833 | -1.127521 |
| C | 6.675508 | -0.022183 | 0.180986 |

```
O    6.796909   1.399933   0.080616
P    6.235297   2.312314  -1.077266
O    7.571068   2.721022  -1.885626
P    8.433327   4.029580  -2.127009
O    9.887806   3.521138  -1.875696
P   11.321475   3.653127  -2.723855
O   11.112631   4.887803  -3.595123
O    7.223982  -2.236468   2.400910
O    5.150003  -1.820654   4.041949
O    5.709644   3.555181  -0.415827
O    5.342659   1.580368  -1.995899
O    8.246764   4.470205  -3.546439
O    8.077043   5.063826  -1.093594
O   12.356882   3.840869  -1.661488
O   11.433976   2.378364  -3.517332
O    9.713503   5.174158   1.404418
O   12.230864   7.350195  -2.251523
O   15.112046   3.708783  -2.466214
O   13.328584   1.659443  -0.057317
O   12.470492   5.863210   0.323821
O    5.642065  -1.003469  -3.450953
O    8.413788  -0.724556  -4.227877
O    2.624781   2.740919  -1.830359
O    5.062460   2.206651  -4.935351
O    3.340549   5.402558  -2.623285
O    3.268873   2.671399   1.078026
O   14.045374   0.981916  -2.870891
O   10.375003  -0.069331  -2.231987
O   13.088559   2.674845  -5.957276
O    9.619907   1.628424  -5.636770
O    8.557662   8.028516  -2.204608
O    5.459659   5.086721  -4.657410
O    9.533903   4.335150  -6.458750
O    9.249968   7.291647  -4.765676
O   11.976710   4.533614  -7.783921
Mg   9.587825   5.319249  -4.716460
Mg   6.399835   5.338064  -0.096349
O    5.042777   6.616770  -0.780055
O    7.032802   5.851096   1.734625
H    3.543754  -3.891299  -1.849023
H   -1.289395  -2.313371  -0.725203
H   -0.861864  -3.318085  -2.021692
H    1.259320  -0.225029   2.837334
H    3.582662  -0.341848   3.161135
H    4.329922  -3.010147   2.592997
H    5.990478  -2.232022   4.213933
H    5.869249  -2.439883   0.874818
H    7.465607  -3.122978   2.162373
H    6.466972   0.161944   2.300684
H    7.675730  -0.428620   0.164149
H    6.112873  -0.405042  -0.654313
H   14.211303   3.891632  -2.192413
H   12.313867   7.044711  -1.351113
H   12.467881   5.120780  -0.283507
H   12.965685   2.442298  -0.474499
H   10.213543   1.828022  -4.907365
```

```
H    12.564175    2.599979   -5.159284
H    13.178979    1.312016   -3.101922
H    10.732207    0.736986   -2.601656
H     3.437697    2.277367   -2.025101
H     5.170884    1.976561   -4.016429
H     8.620939    0.023945   -4.782392
H     5.471818   -0.206892   -2.955973
H     4.028637    3.174520    0.798353
H     3.038883    4.528262   -2.358127
H     9.487713    5.192485    0.477869
H     8.529655    7.184312   -1.763699
H     8.955037    7.691732   -3.933781
H    12.446954    3.895501   -7.239905
H     6.307144    5.015089   -4.227297
H     9.440902    3.384160   -6.328654
H     6.552660   -0.929552   -3.735458
H     8.989790   -0.628177   -3.469743
H     2.572828    2.726203   -0.876651
H     5.162073    3.156136   -4.966635
H    10.462553    0.009721   -1.282196
H    14.604890    1.755583   -2.881667
H    10.141847    1.149180   -6.271569
H    13.988965    2.777371   -5.671468
H    11.684828    5.755331    0.852784
H    11.896260    6.599747   -2.737010
H    13.694439    1.167224   -0.788328
H    15.306488    4.320078   -3.166784
H     8.822555    7.747197   -5.484681
H    12.066454    4.244227   -8.685332
H     9.683682    4.246291    1.639857
H     7.708403    8.437288   -2.042735
H     3.606477    1.790550    1.217718
H     2.575596    5.867931   -2.943690
H     4.816272    5.247538   -3.969252
H    10.296580    4.452803   -7.038427
H     5.273276    7.512535   -1.046483
H     4.390073    6.276122   -1.404727
H     6.696727    6.567479    2.265073
H     7.978734    5.761306    1.899928
O     5.930027    9.173328   -1.580384
O    12.771620    7.289343   -7.199866
H    13.467870    7.260605   -6.554727
H    12.555961    6.378697   -7.393429
H     5.959206    9.830981   -0.892722
H     5.463091    9.575138   -2.306195
H     8.347178    2.152908    1.353137
O     9.155302    2.323084    1.830587
H     8.973963    2.096378    2.736471
H    10.342063    0.926161    1.027361
O    10.902707    0.275293    0.610465
H    11.746330    0.708590    0.473458
```

**Mg-ADP⁻**

| | | | |
|---|---:|---:|---:|
| C | -5.727742 | -2.073887 | 1.617346 |
| C | -5.224173 | -0.847915 | 2.073160 |
| C | -3.876909 | -0.631798 | 1.896894 |
| N | -3.016721 | -1.479160 | 1.334261 |
| C | -3.598978 | -2.591996 | 0.947199 |
| N | -4.878976 | -2.931823 | 1.052421 |
| N | -5.814779 | 0.239375 | 2.692804 |
| C | -4.846308 | 1.062750 | 2.872635 |
| N | -3.633790 | 0.606724 | 2.423419 |
| C | -2.398543 | 1.350543 | 2.474325 |
| O | -2.010089 | 1.691152 | 1.168649 |
| C | -0.609985 | 1.548361 | 0.998441 |
| C | -0.243573 | 0.406305 | 1.939616 |
| C | -1.206082 | 0.618738 | 3.116842 |
| O | -0.664460 | 1.490885 | 4.070920 |
| O | 1.105361 | 0.481168 | 2.315691 |
| C | -0.317268 | 1.278175 | -0.457548 |
| O | -0.732403 | 2.412986 | -1.194019 |
| P | -0.792454 | 2.451343 | -2.787461 |
| O | 0.721488 | 2.585087 | -3.244411 |
| P | 1.815139 | 3.827202 | -3.145187 |
| O | 2.562101 | 3.618876 | -1.859478 |
| N | -7.015103 | -2.431326 | 1.759827 |
| O | -1.305224 | 1.160342 | -3.319535 |
| O | -1.550783 | 3.695474 | -3.123086 |
| Mg | -0.932277 | 5.604549 | -3.398920 |
| O | -0.834704 | 5.411240 | -5.522421 |
| O | 2.652811 | 3.639048 | -4.392242 |
| O | 0.986312 | 5.093620 | -3.176283 |
| O | -2.935261 | 6.331387 | -3.758045 |
| O | -1.333555 | 6.005480 | -1.337051 |
| O | 2.974333 | 1.116462 | -5.571482 |
| O | 1.822089 | 5.287023 | -6.505766 |
| O | -3.082198 | 4.305478 | 0.130031 |
| O | -1.575569 | 0.602299 | -6.088784 |
| O | 2.262469 | 4.856680 | 0.669446 |
| O | 5.387176 | 3.420986 | -1.693213 |
| O | 0.886826 | 1.083143 | -7.554150 |
| O | 1.365529 | 3.609508 | -8.804159 |
| O | 2.937550 | 2.761254 | 2.641401 |
| O | 1.830780 | 7.957636 | -5.380714 |
| O | 5.410157 | 4.185569 | -4.553409 |
| H | -2.964359 | -3.324531 | 0.483245 |
| H | -7.690875 | -1.732284 | 1.978716 |
| H | -7.342371 | -3.220232 | 1.245308 |
| H | -4.932558 | 2.026979 | 3.330946 |
| H | -2.607391 | 2.243619 | 3.042705 |
| H | -1.501858 | -0.310888 | 3.582916 |
| H | 0.225162 | 1.220208 | 4.269619 |
| H | -0.452987 | -0.543973 | 1.471042 |
| H | 1.466462 | -0.387893 | 2.444381 |
| H | -0.112602 | 2.459946 | 1.310587 |
| H | 0.745258 | 1.116502 | -0.590825 |
| H | -0.858241 | 0.404355 | -0.792564 |
| H | 0.063304 | 0.948062 | -7.084748 |
| H | 1.579346 | 1.039109 | -6.893777 |
| H | 3.848566 | 1.100179 | -5.942343 |
| H | 2.878904 | 1.974455 | -5.143496 |
| H | -3.226296 | 6.434420 | -4.664989 |
| H | -3.643079 | 5.904199 | -3.287566 |
| H | -1.489096 | 6.894436 | -1.038894 |
| H | -1.924410 | 5.432999 | -0.839757 |
| H | -1.485736 | 0.792121 | -5.150847 |
| H | -1.834734 | -0.309723 | -6.148185 |
| H | 0.034192 | 5.413381 | -5.930931 |
| H | -1.443654 | 5.856499 | -6.107496 |
| H | 2.147652 | 4.696589 | -5.816533 |
| H | 1.759576 | 4.768175 | -7.308811 |
| H | 2.661166 | 8.121305 | -4.948353 |
| H | 1.925309 | 7.115143 | -5.825646 |
| H | 5.787328 | 3.545866 | -5.145463 |
| H | 4.463635 | 4.010942 | -4.537550 |
| H | 5.628073 | 3.657913 | -2.585416 |
| H | 4.427518 | 3.463099 | -1.688997 |
| H | 2.329955 | 2.029760 | 2.551140 |
| H | 2.812203 | 3.096705 | 3.521531 |
| H | 2.072884 | 3.469590 | -9.422863 |
| H | 1.168509 | 2.749229 | -8.425246 |
| H | 2.483685 | 4.169770 | 1.294233 |
| H | -3.437200 | 4.662849 | 0.935819 |
| H | -2.696640 | 3.463287 | 0.363982 |
| H | 2.326216 | 4.455964 | -0.200859 |
| O | -0.329561 | 7.655164 | -3.551831 |
| H | 0.414535 | 7.800801 | -4.144194 |
| H | -0.995957 | 8.289667 | -3.792203 |
| H | -2.879153 | 0.258163 | -2.774953 |
| O | -3.674661 | -0.220093 | -2.532703 |
| H | -3.615523 | -0.361552 | -1.595523 |
| O | -3.203709 | 6.676603 | -6.587334 |
| H | -3.800074 | 6.159670 | -7.119010 |
| H | -3.249495 | 7.563521 | -6.929016 |

**ATP$^{4-}$ (for valence-ionization calculations)**
| | | | |
|---|---|---|---|
| C | -2.400558 | 0.703779 | 5.179742 |
| C | -1.639049 | 1.418460 | 4.047763 |
| O | -1.537477 | 0.507256 | 2.965714 |
| C | -1.850156 | -0.815965 | 3.408783 |
| C | -2.873870 | -0.608153 | 4.523536 |
| N | -2.254726 | 2.638068 | 3.567285 |
| C | -3.551119 | 2.834866 | 3.142775 |
| C | -3.606353 | 4.157934 | 2.719384 |
| N | -2.368910 | 4.762856 | 2.866975 |
| C | -1.604144 | 3.827684 | 3.364102 |
| N | -4.562770 | 1.960829 | 3.129693 |
| C | -5.677445 | 2.495980 | 2.644561 |
| N | -5.868479 | 3.746405 | 2.202622 |
| C | -4.844856 | 4.611427 | 2.231082 |
| N | -5.063939 | 5.898288 | 1.837632 |
| C | -2.372804 | -1.627764 | 2.252020 |
| O | -1.303659 | -1.870314 | 1.338756 |
| P | -1.603465 | -2.638537 | -0.065133 |
| O | -0.257493 | -2.755909 | -0.745376 |
| O | -2.833756 | -1.695303 | 5.430226 |
| O | -1.483809 | 0.426406 | 6.222334 |
| O | -2.144292 | -4.097608 | 0.429502 |
| P | -1.488680 | -5.492236 | 1.012285 |
| O | -0.808102 | -6.216751 | -0.138368 |
| O | -2.753772 | -1.997422 | -0.786583 |
| O | -0.416924 | -4.902817 | 2.048588 |
| P | 0.911495 | -5.648371 | 2.839531 |
| O | 0.986696 | -4.866908 | 4.145223 |
| O | -2.616575 | -6.200409 | 1.713097 |
| O | 0.525045 | -7.126439 | 2.992494 |
| O | 2.085018 | -5.417255 | 1.877166 |
| O | -2.440103 | 7.244319 | 1.553407 |
| O | -5.948830 | 7.463662 | 4.205323 |
| H | -6.544673 | 1.843386 | 2.596606 |
| H | -4.248402 | 6.469177 | 1.598172 |
| H | -5.879416 | 6.038149 | 1.251217 |
| H | -0.555659 | 3.940837 | 3.607057 |
| H | -0.646174 | 1.697740 | 4.409200 |
| H | -3.230830 | 1.307747 | 5.556052 |
| H | -1.707691 | -0.462083 | 6.559356 |
| H | -3.874512 | -0.475854 | 4.106798 |
| H | -3.711388 | -1.828173 | 5.820013 |
| H | -0.952044 | -1.294049 | 3.821868 |
| H | -2.758070 | -2.579979 | 2.633497 |
| H | -3.181540 | -1.086968 | 1.749593 |
| H | -5.674149 | 6.923880 | 3.435035 |
| H | -6.752811 | 7.038134 | 4.539461 |
| H | -2.181024 | 6.465249 | 2.100235 |
| H | -2.354271 | 8.017895 | 2.130740 |
| H | 0.042733 | -4.072017 | -1.989039 |
| O | 0.109469 | -4.959778 | -2.401915 |
| H | -0.194443 | -5.529753 | -1.658108 |
| H | -0.089433 | -7.883736 | 0.224392 |
| O | 0.351080 | -8.605645 | 0.722261 |
| H | 0.493711 | -8.158918 | 1.597107 |
| H | 2.176709 | -3.957348 | 0.970348 |
| O | 2.248297 | -3.133638 | 0.420393 |
| H | 1.352838 | -3.000717 | 0.040389 |

**[Mg-ATP]$^{2-}$ (for valence-ionization calculations)**

| | | | |
|---|---|---|---|
| C | -4.047292 | -1.071794 | 4.817318 |
| C | -3.194817 | -0.114970 | 4.221151 |
| C | -2.475715 | -0.525112 | 3.108038 |
| N | -2.513154 | -1.740345 | 2.542950 |
| C | -3.346534 | -2.554577 | 3.181839 |
| N | -4.097823 | -2.297510 | 4.259171 |
| N | -2.906296 | 1.205666 | 4.524314 |
| C | -2.038933 | 1.568291 | 3.617205 |
| N | -1.725954 | 0.569169 | 2.731112 |
| C | -0.823626 | 0.697754 | 1.608148 |
| O | -1.583590 | 0.680709 | 0.409691 |
| C | -0.873654 | -0.046794 | -0.596203 |
| C | -0.141105 | -1.136025 | 0.183624 |
| C | 0.238009 | -0.414038 | 1.491974 |

S29

| | | | |
|---|---|---|---|
| O | 1.509635 | 0.195558 | 1.365199 |
| O | 1.001912 | -1.561784 | -0.537184 |
| C | -1.849543 | -0.577298 | -1.618845 |
| O | -2.425337 | 0.486500 | -2.385432 |
| P | -1.934129 | 0.768888 | -3.907146 |
| O | -0.384407 | 1.198119 | -3.735908 |
| P | 0.427766 | 2.616810 | -3.422080 |
| O | -0.545610 | 3.273680 | -2.307929 |
| P | -1.304772 | 4.770621 | -2.201058 |
| O | -0.228084 | 5.832729 | -2.181599 |
| N | -4.796907 | -0.813912 | 5.897814 |
| O | -1.961991 | -0.481325 | -4.730657 |
| O | -2.775822 | 1.950626 | -4.359405 |
| Mg | -3.725604 | 3.574997 | -3.794829 |
| O | -5.503043 | 2.565295 | -4.360801 |
| O | -4.328064 | 4.919987 | -5.224733 |
| O | -2.242566 | 4.829846 | -3.433769 |
| O | -4.279909 | 3.525557 | -1.819404 |
| O | 1.703808 | 2.182651 | -2.754562 |
| O | 0.476035 | 3.424755 | -4.688084 |
| O | -2.125922 | 4.610244 | -0.908624 |
| H | -3.433038 | -3.561673 | 2.782199 |
| H | -4.769928 | 0.096850 | 6.362568 |
| H | -5.378617 | -1.549417 | 6.274898 |
| H | -1.585350 | 2.547691 | 3.539785 |
| H | -0.316762 | 1.659137 | 1.726105 |
| H | 0.224317 | -1.080904 | 2.358273 |
| H | 1.997752 | -0.316365 | 0.692341 |
| H | -0.808943 | -1.973247 | 0.398676 |
| H | 1.203104 | -2.484079 | -0.316149 |
| H | -0.139566 | 0.606418 | -1.083368 |
| H | -1.339605 | -1.290972 | -2.273015 |
| H | -2.673594 | -1.089352 | -1.115152 |
| H | -5.341400 | 1.607452 | -4.389346 |
| H | -6.343071 | 2.685749 | -3.888916 |
| H | -5.180365 | 4.968544 | -5.684459 |
| H | -3.927563 | 5.803833 | -5.234110 |
| H | -5.115404 | 3.895406 | -1.496983 |
| H | -3.509291 | 3.976743 | -1.318765 |
| H | -3.013344 | 1.480029 | 9.793725 |
| O | -2.922637 | 2.235135 | 9.191427 |
| H | -3.194565 | 3.008622 | 9.710031 |
| H | -3.780065 | 1.923758 | 6.074274 |
| O | -4.410303 | 1.929844 | 6.828719 |
| H | -3.873395 | 2.033819 | 7.643467 |

# References

1. ACD/ChemSketch. Advanced Chemistry Development, Inc. (ACD Labs): Toronto, ON, Canada.
2. Storer, A.C. and A. Cornish-Bowden, *Concentration of MgATP2- and other ions in solution. Calculation of the true concentrations of species present in mixtures of associating ions.* Biochem J, 1976. **159**(1): p. 1-5.
3. *Maple 2016*. Maplesoft, a division of Waterloo Maple Inc.: Waterloo, Ontario.
4. Bock, J.L., et al., *25Mg NMR Studies of magnesium binding to erythrocyte constituents.* Journal of Inorganic Biochemistry, 1991. **44**(2): p. 79-87.
5. Winter, B., et al., *Full Valence Band Photoemission from Liquid Water Using EUV Synchrotron Radiation.* The Journal of Physical Chemistry A, 2004. **108**(14): p. 2625-2632.
6. Weber, R., et al., *Photoemission from Aqueous Alkali-Metal−Iodide Salt Solutions Using EUV Synchrotron Radiation.* The Journal of Physical Chemistry, 2004. **108**: p. 4729-4736.
7. Schroeder, C., *Quantifying Aspects of DNA Damage. Ph.D. dissertation*. University of Southern California: Los Angeles, CA USA
8. Schroeder, C.A., et al., *Oxidation Half-Reaction of Aqueous Nucleosides and Nucleotides via Photoelectron Spectroscopy Augmented by ab Initio Calculations.* Journal of the American Chemical Society, 2015. **137**(1): p. 201-209.
9. Frańska, M., et al., *Gas-Phase Internal Ribose Residue Loss from Mg-ATP and Mg-ADP Complexes: Experimental and Theoretical Evidence for Phosphate-Mg-Adenine Interaction.* Journal of the American Society for Mass Spectrometry, 2022. **33**(8): p. 1474-1479.